\documentclass[epj]{svjour}

\usepackage{graphicx,amsmath,amssymb,bm,xcolor,tabularx}
\usepackage{enumerate}
\usepackage{multirow,dcolumn}
\usepackage{color}
\usepackage{isotope}
\usepackage{hhline}
\usepackage{bbm}

\newcommand{\mev}{\,\mathrm{MeV}}

\newcommand{\fmiq}{\,\mathrm{fm}^{-3}}

\newcommand{\km}{\,\mathrm{km}}

\newcommand{\beq}{\begin{equation}}
\newcommand{\eeq}{\end{equation}}
\newcommand{\beqa}{\begin{eqnarray}}
\newcommand{\eeqa}{\end{eqnarray}}

\newcommand{\nxlo}[1]{%
   \ifnum0=#1\relax%
      \text{LO}%
   \else%
   \ifnum1=#1\relax%
      \text{NLO}%
   \else%
      \text{N}\ensuremath{^{#1}}\text{LO}%
   \fi
   \fi
}

\begin{document}
\title{Confronting gravitational-wave observations with modern nuclear physics constraints}
\author{I.\ Tews\inst{1} \and J.\ Margueron\inst{2} \and S.\ Reddy\inst{3,4}}                     
\institute{Theoretical Division, Los Alamos National Laboratory, Los Alamos, NM 87545, USA 
\and Institut de Physique Nucl\'eaire de Lyon, CNRS/IN2P3, Universit\'e de Lyon, Universit\'e Claude Bernard Lyon 1, F-69622 Villeurbanne Cedex, France  
\and Institute for Nuclear Theory, University of Washington, Seattle, WA 98195-1550, USA
\and JINA-CEE, Michigan State University, East Lansing, MI, 48823, USA }
\date{Received: date / Revised version: date}
%
\abstract{Multi-messenger observations of neutron star (NS) mergers have the potential to revolutionize nuclear astrophysics. They will improve our understanding of nucleosynthesis, provide insights about the equation of state (EOS) of strongly-interacting matter at high densities, and enable tests of the theory of gravity and of dark matter. Here, we focus on the EOS, where both gravitational waves (GWs) from neutron-star mergers and X-ray observations from space-based detectors such as NICER will provide more stringent constraints on the structure of neutron stars. Furthermore, recent advances in nuclear theory have enabled reliable calculations of the EOS at low densities using effective field theory based Hamiltonians and advanced techniques to solve the quantum many-body problem. In this paper, we address how the first observation of GWs from GW170817 can be combined with modern calculations of the EOS to extract useful insights about the EOS of matter encountered inside neutron stars. We analyze the impact of various uncertainties, the role of phase transitions in the NS core, and discuss how future  observations will improve our understanding of dense matter. 
} 
\maketitle

\section{Introduction}
\label{sec:intro}

Multimessenger observations of neutron-star (NS) mergers have the potential to revolutionize nuclear astrophysics much in the same way as observations of the cosmic microwave background (CMB) radiation revolutionized particle astrophysics. Neutron-star merger events simultaneously emit gravitational waves (GWs) and electromagnetic (EM) signals, from gamma-rays, X-rays, optical, infrared, to radio waves, and neutrinos. The first observation of a NS merger, GW170817 in the GW spectrum, GRB 170817A in the gamma-ray spectrum, and AT~2017gfo in the electromagnetic (EM) spectrum, was made on August 17, 2017, and in the weeks thereafter~\cite{TheLIGOScientific:2017qsa,GBM:2017lvd,Monitor:2017mdv,Abbott:2018wiz}. Triggered by the Fermi and Integral telescopes~\cite{Monitor:2017mdv,Savchenko:2017ffs}, this observation provided detailed spectral and temporal features both in GWs and EM radiation. Theoretical efforts to interpret this data has provided insights into the production of heavy r-process elements in NS mergers~\cite{Drout:2017ijr}, and constraints on the EOS of dense matter~\cite{Annala:2017llu,Fattoyev:2017jql,Most:2018hfd,Lim:2018bkq,Tews:2018iwm}. NS mergers have the potential to provide detailed information on the properties of the merging compact stars, such as their masses and radii~\cite{Bauswein:2017vtn}, as well as on the properties of the densest baryonic matter to be observed in the universe. Future detections of NS mergers, anticipated during the next observing run of the Advanced LIGO and VIRGO detectors, could provide even stronger constraints on the EOS of strongly-interacting matter and the r-process. 

We are pleased to contribute to this topical issue on "First joint gravitational wave and electromagnetic observations:  Implications for nuclear physics", which contains several articles devoted to the theory and computing needed to improve the description of dense matter and to model neutron-star mergers - efforts that will play a key role in extracting insights from GW170817 and future detections. Here, we elaborate on earlier work in Ref.~\cite{Tews:2018iwm}, where we analyzed GW170817 constraints on the dense matter EOS, to provide additional details, discussions, and new results. 

Our contribution is structured as follows. In Sec.~\ref{sec:models} we describe the NS equation-of-state models employed in our analysis. In particular, we use two models: the minimal model or meta-model (MM), see Sec.~\ref{sec:minmod} and the maximal or speed-of-sound model (CSM), see Sec.~\ref{sec:maxmod}. Both models are constrained at low densities by state-of-the-art calculations of neutron-rich matter from chiral effective field theory (EFT). We discuss these models in the context of GW170817 in great detail in Sec.~\ref{sec:results} and  analyze the impact of phase transitions or future GW detections. Finally, we summarize our results and provide an outlook in Sec.~\ref{sec:summary}.

\section{Models}
\label{sec:models}

In this section, we discuss the dense-matter models we use in our analysis. Calculations of the EOS of neutron matter based on Hamiltonians derived from chiral EFT provide a reliable method to estimate the uncertainties associated with poorly constrained aspects of two- and many-body nuclear forces at short-distance~\cite{Lynn:2015jua,Tews:2018kmu}. Chiral EFT is a systematic expansion for nuclear forces in powers of momenta, and provides an efficient way to estimate theoretical uncertainties. It is however limited to momenta up to the so-called breakdown scale, $\Lambda_b$, which signals the breakdown of the effective theory due to additional high-momentum physics, e.g. the onset of new degrees of freedom. Since $\Lambda_b$ is expected to be of the order of $\simeq 500-600$~MeV~\cite{Melendez:2017phj}, chiral EFT is not applicable at all densities encountered in neutron stars and chiral EFT interactions have typically been used to describe neutron matter only up to saturation density, $n_{sat}$.  Here, using insights obtained in Ref.~\cite{Tews:2018iwm}, we will analyze to which extent chiral EFT predictions up to 2$n_{sat}$ with conservative error estimates provide useful constrains for the nuclear equation of state, even though uncertainties grow fast with density.

To describe the EOS at higher densities, we will consider two extrapolation schemes rooted in low-density microscopic predictions and widely covering our present uncertainties at higher density. These two schemes are the minimal model or meta-model (MM), based on a smooth extrapolation of chiral EFT results, and the maximal model or speed-of-sound model (CSM), which explores the widest possible domain for the EOS and contains also more drastic behavior with density; see Ref.~\cite{Tews:2018iwm} for the first analysis of GWs with these models. These two models show some overlap for properties of dense neutron-star matter, as suggested from the masquerade phenomenon~\cite{Alford:2004pf}, but also highlight differences: The confrontation of these models with each other and with observations sheds light on the impact of the presence of strong phase transitions at high density, as is detailed hereafter.

\subsection{Pure neutron matter from chiral EFT}

\begin{figure*}[t]
\begin{center}
\includegraphics[trim= 0.0cm 0 2.0cm 1.5cm, clip=,width=0.9\columnwidth]{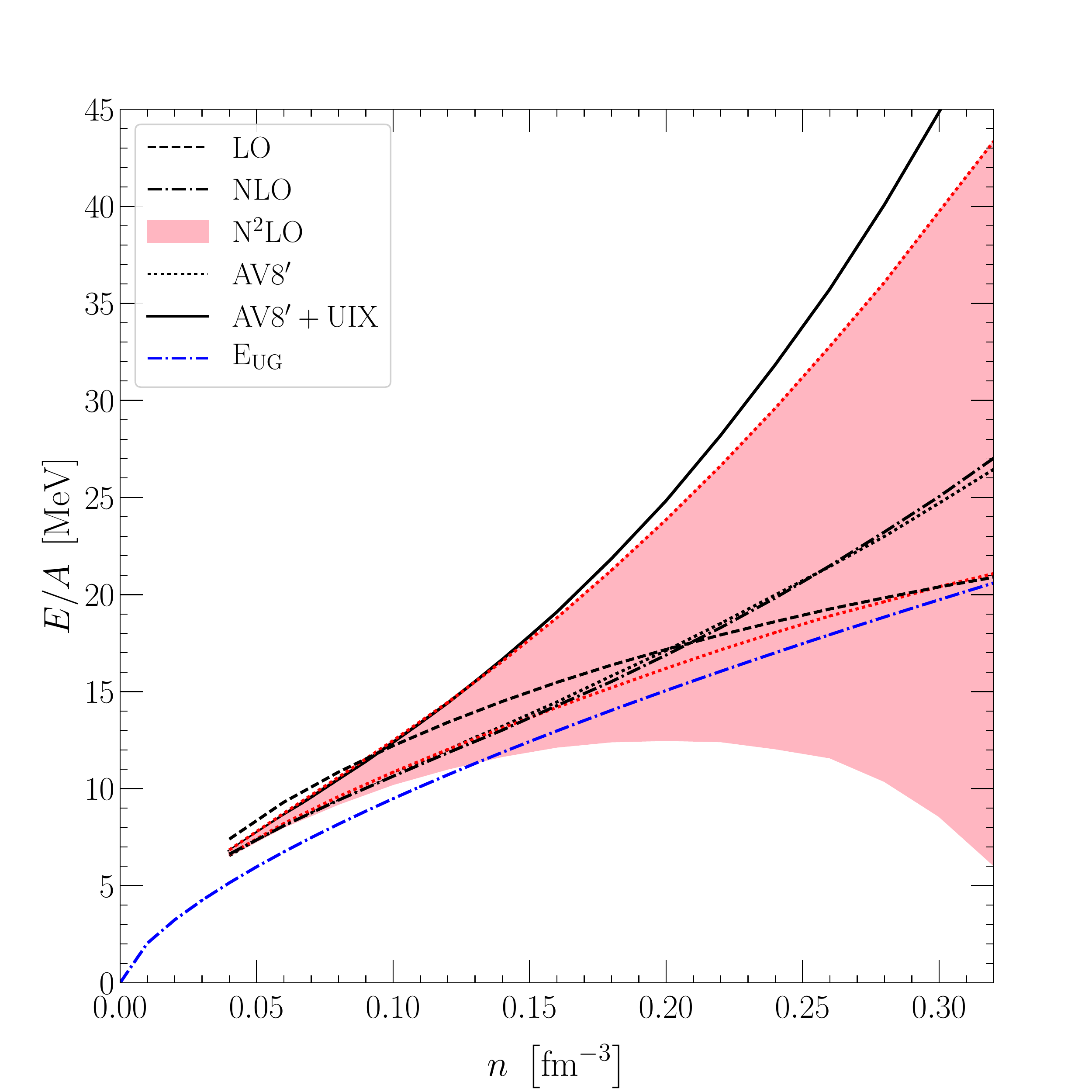}\hspace{0.4cm}
\includegraphics[trim= 0.0cm 0 2.0cm 1.5cm,
clip=,width=0.9\columnwidth]{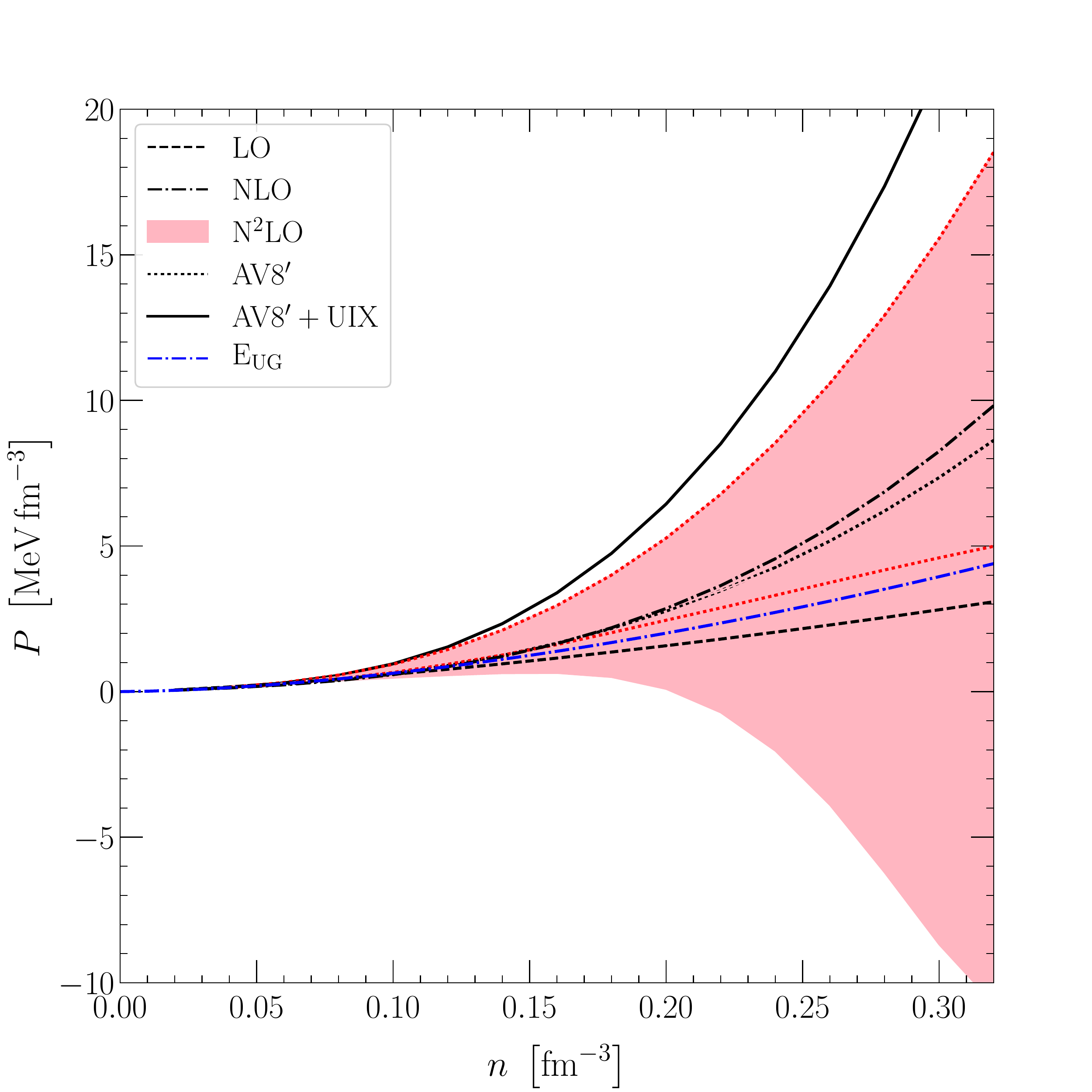}
\end{center}
\caption{\label{fig:chiralPNM} The energy per particle and pressure of pure neutron matter as functions of baryon density up to $2n_{\rm sat}$. We show the constraints from Ref.~\cite{Tews:2018kmu} based on AFDMC calculations with local chiral potentials at N$^2$LO (red bands). As a comparison, we show results at LO (black dashed lines), NLO (black dashed-dotted lines), as well as calculations using phenomenological $NN$ interactions only (black dotted lines) and including also phenomenological $3N$ forces (black solid lines). We also indicate the unitary-gas bound of Ref.~\cite{Kolomeitsev:2016sjl} (blue dashed-dotted lines) and the part of the uncertainty band that we use for our NS modeling (red dotted lines); see text for more details.}
\end{figure*} 

Neutron stars are ideal laboratories to test theories of the strong interaction at finite chemical potential: the structure of neutron stars is governed by the knowledge of the EOS of neutron-star matter, relating energy density, pressure, and temperature. Additional uncertainties may come from rotation and magnetic field distribution in the star, but the dense-matter EOS is the key input. Since neutron stars explore densities from a few gram per cubic centimeter up to 10 times the nuclear saturation density, $n_{\rm sat}=0.16 \fmiq = 2.7\!\cdot\! 10^{14} \rm{g\, cm}^{-3}$, the knowledge of the EOS is required for densities covering several orders of magnitude. Though young proto-neutron stars or neutron-star remnants also explore the EOS at high temperatures up to several tens of MeV, older neutron stars can typically be considered as cold objects at $T=0$. This is especially true for two binary NS during the inspiral phase of a neutron-star merger, whose properties can be analyzed from the premerger GW signal.  

While the EOS of the neutron-star crust, reaching up to $n_{\rm sat}/2$, is rather well constrained, the uncertainty of the EOS increases fast with density and the composition of the inner core of NS is still unknown. Nevertheless, in the density range from $n_{\rm sat}/2$ up to about $2n_{\rm sat}$, the neutron-star EOS can be constrained by state-of-the-art nuclear-theory models. The starting point for these constraints are calculations of pure neutron matter (PNM). PNM is an idealized, infinite system consisting solely of neutrons, but it is much easier to compute than systems containing also protons. The reason is that certain parts of the nuclear interaction, e.g., tensor interactions, are weaker or do not contribute at all among neutrons. In contrast to symmetric nuclear matter, PNM is also not unstable with respect to density fluctuations below $n_\mathrm{sat}$, and uniform matter remains the true ground state of PNM at all densities, simplifying its calculation.

To reliably describe neutron matter, one needs precise and accurate quantum many-body methods in combination with a reliable model for the nuclear interaction. Neutron matter has been extensively studied in the last decade, using a multitude of nuclear interactions and advanced \textsl{ab initio}  many-body methods. Among these are, e.g., many-body perturbation theory~\cite{Hebeler:2009iv,Drischler:2016djf,Holt:2016pjb}, the coupled-cluster method~\cite{Hagen:2013yba}, quantum Monte Carlo methods~\cite{Gandolfi:2011xu}, or the self-consistent Green's function method~\cite{Carbone:2014mja}. A comparison of these different studies, see e.g., Refs.~\cite{Gandolfi:2015jma,Hebeler:2015hla}, shows that neutron matter is rather well constrained by these multiple \textsl{ab initio} approaches using diverse nuclear Hamiltonians.  In this paper, we will use calculations of neutron matter obtained with the auxiliary-field diffusion Monte Carlo (AFDMC) method~\cite{Carlson:2014vla} together with modern nuclear Hamiltonians from chiral EFT. 

Quantum Monte Carlo methods are among the most precise many-body methods for strongly interacting systems~\cite{Carlson:2014vla}. They provide the ground state of a many-body system, governed by a non-relativistic nuclear Hamiltonian defining the Schr\"odinger equation, by evolving a trial wave function $\Psi_T$ in imaginary time, 
\begin{equation}
\Psi_{GS}=\lim_{\tau \to \infty} e^{- \mathcal{H}\tau}\Psi_T\,,
\end{equation}
where $\Psi_T$ is constructed so that it has a non-vanishing overlap with the ground state $\Psi_{GS}$. Expanding $\Psi_T$ in eigenfunctions of the Hamiltonian, one can easily see that contributions of excited states decay with time, and only the ground-state component of the trial wave function remains. Quantum Monte Carlo methods have been used to successfully describe nuclei up to \isotope[16]{O}~\cite{Carlson:2014vla,Piarulli:2017dwd,Lonardoni:2017hgs} and neutron matter~\cite{Gandolfi:2011xu,Lynn:2015jua}. At very low densities, where neutron matter is close to the unitary limit and interactions are dominated by large scattering-length physics, these methods~\cite{Carlson:2008zza} have been successfully confronted to experimental measurements of cold atomic gases~\cite{Nascimbene2010,Navon2010,Zwierlein:2015}. Due to its great success to study strongly-interacting matter and nuclei~\cite{Gandolfi:2011xu,Lonardoni:2014bwa,Lynn:2015jua,Gandolfi:2016bth,Lonardoni:2017hgs}, we employ in this work the AFDMC method to determine PNM properties. For more details on Quantum Monte Carlo methods we refer the reader to Ref.~\cite{Carlson:2014vla}. 

On the interaction side, chiral EFT~\cite{Epelbaum2009,Machleidt:2011zz} is a modern theory for nuclear forces that is consistent with the symmetries of Quantum Chromodynamics and systematically describes the nucleon-nucleon interaction in terms of explicitly resolved longer-range pion exchanges as well as short-range nucleon contact interactions.  Chiral EFT is based on a momentum expansion in terms of $p/\Lambda_b$, where $p$ is the typical momentum of the nuclear system at hand, and $\Lambda_b$ is the breakdown scale already discussed. The short-range interaction terms parametrize all unresolved and unknown high-energy physics beyond the breakdown scale, and depend on a set of low-energy couplings (LECs), which are typically fitted to nucleon-nucleon ($NN$) scattering data and properties of light nuclei. Chiral EFT does not only describe $NN$ interactions but also consistent three-body ($3N$) and higher many-body forces. It has been successfully applied to calculate properties of ground and excited states of nuclei, nuclear matter, as well as electroweak processes; see, e.g, Ref.~\cite{Hebeler:2015hla} for a review. Most importantly, the systematic chiral EFT expansion enables the estimation of theoretical uncertainties for these physical systems.

In our analysis in this work, we use local chiral EFT interactions that have been constructed especially for the use in QMC methods in Refs.~\cite{Lynn:2015jua,Gezerlis:2013ipa,Gezerlis:2014zia,Tews:2015ufa}. These interactions have been successfully tested in light- to medium-mass nuclei and in n-$\alpha$ scattering~\cite{Lynn:2015jua,Lonardoni:2017hgs} and agree with our current knowledge of the empirical parameters of nuclear matter~\cite{Kolomeitsev:2016sjl,Margueron:2017eqc}. In Ref.~\cite{Tews:2018kmu}, these interactions have been used to study neutron matter up to $2n_{\rm sat}$ with theoretical uncertainty estimates using the AFDMC method.  
For more details on QMC calculations with local chiral interactions we refer the reader to Ref.~\cite{Lynn:2019rdt}. 

In particular, in this work we use local chiral interactions at a cutoff scale $R_0=1.0$ fm with its systematic uncertainty estimates. In Fig.~\ref{fig:chiralPNM} we show the results for the energy per particle and pressure of neutron matter at leading order (LO), next-to-leading order (NLO), and at next-to-next-to-leading order (N$^2$LO) with its uncertainty band for densities ranging from 0.04~fm$^{-3}$  up to $2n_{\rm sat}$. We find that the uncertainty bands increase fast with density and are quite sizable at $2 n_{\rm sat}$. In addition to the results for chiral interactions, we also show in Fig.~\ref{fig:chiralPNM} AFDMC results employing the phenomenological AV8' $NN$ and AV8' $NN$ plus UIX $3N$ interactions as a comparison. It is interesting to note that the AV8' and NLO $NN$ interactions agree very well with each other, which highlights the fact that many-body forces are a considerable source of uncertainty. Finally, we also compare all calculations with the unitary-gas limit of Ref.~\cite{Kolomeitsev:2016sjl}.

\subsection{Discussion of uncertainties}

The uncertainty bands shown in Fig.~\ref{fig:chiralPNM} include the following sources of uncertainty: i) the truncation of the nuclear Hamiltonian within the chiral expansion, ii) the regularization scheme and scale, which are needed to implement nuclear Hamiltonians in many-body methods, iii) the uncertainties in the determination of low-energy couplings from data, and iv) the many-body uncertainty that originates in approximations made when solving the Schr\"odinger equation for the nuclear many-body system. The first three sources, which originate in the nuclear Hamiltonian, dominate over the many-body uncertainty from QMC methods. Among these three, the truncation uncertainty is the dominant source of uncertainty and we will discuss it in the following.

The truncation uncertainty can be expressed in the following way. 
Introducing the dimensionless expansion parameter $Q=p/\Lambda_b$ and following Ref.~\cite{Furnstahl:2015rha}, under the prerequisite that chiral EFT is a converging theory, one can define the order-by-order contributions to an observable $X$ using the following infinite summation,
\begin{equation}\label{eq:chiralExp}
X=X_0\sum_{i=0}^{\infty} c_i Q^{i}\,.
\end{equation}
Here, $X_0$ sets the natural scale expected for the observable $X$, e.g., the leading-order result, $X_0=X_{\rm{LO}}$ ($c_0=1$), and the $c_{i\ge 1}$ denote the expansion coefficients. In calculations of nuclear systems, due to practical reasons this sum has to be truncated at a certain order $n$, inducing the so-called truncation uncertainty. This uncertainty is intrinsic to \emph{all} nuclear Hamiltonians but can be specified for chiral EFT Hamiltonians by  
\begin{equation}
\Delta X=X- X_0\sum_{i=0}^{n}c_i Q^{i}\,.
\end{equation}

It has been shown in Ref.~\cite{Furnstahl:2015rha} that for practical purposes an estimate of the magnitude of the first truncated term in Eq.~\eqref{eq:chiralExp}, given by $i=n+1$, is a sufficient uncertainty estimate. To obtain this estimate, both the size of the unknown expansion coefficient $c_{n+1}$ and of the expansion parameter $Q$ are required. A conservative choice for the coefficient $c_{n+1}$ is the maximum of all previously found coefficients, 
\begin{equation}
c_{n+1}=\max_{i=0}^n{c_i}\,,
\end{equation} 
while $Q$ has to be estimated from the typical momentum scale for the system at hand. This uncertainty prescription is similar to the one presented by Epelbaum, Krebs, and Mei{\ss}ner (EKM)~\cite{Epelbaum:2014efa}, and the truncation uncertainty, e.g., at N$^2$LO, can be obtained from an order-by-order calculation as 
\begin{align}
\Delta X^{\nxlo{2}}=\max &
\left(\vphantom{X^{\nxlo{2}}}Q^{4} \left|X^{\nxlo{0}}-X^{\rm free}\right|,Q^2 \left|X^{\nxlo{1}}-X^{\nxlo{0}}\right|,\right. \nonumber \\
&\quad \left. Q\left|X^{\nxlo{2}}-X^{\nxlo{1}}\right|
\right)\nonumber\\
&= Q^4 X_0 \max_{i=0}^n{c_i}\,. \label{eq:uncertainty}
\end{align}
We have used this uncertainty estimate to compute the truncation uncertainty, using $Q=\sqrt{3/5}k_F/\Lambda_b$, with the Fermi momentum $k_F$ and $\Lambda_b=500 \mev$. 

\begin{table*}[t]
\centering
\setlength{\tabcolsep}{10pt}
\renewcommand{\arraystretch}{1.2}
\begin{tabular}{cccccccccccc}
\hline
$P_{\alpha}$ & $E_{sat}$ & $E_{sym}$ & $n_{sat}$ & $L_{sym}$ & $K_{sat}$ & $K_{sym}$ & $Q_{sat}$ & $Q_{sym}$ & $Z_{sat}$ & $Z_{sym}$ & $b$\\
                    & MeV           & MeV           & fm$^{-3}$           & MeV           & MeV           & MeV           & MeV           & MeV           & MeV           & MeV & \\
\hline
Max             & -15 & 38 & 0.17 & 90  &  270  &  200  &   1000 &   2000 &   3000 &    3000 & 14 \\
Min              & -17 & 26 & 0.15 & 20  &  190  & -400  &  -1000 &   -2000 &  -3000 & -3000 &1 \\
\hline
\end{tabular}
\caption{Empirical parameters and their domain of variation entering into the definition of the MM~(\ref{eq:MM:energy}). The parameters $\kappa_{sat}$ and $\kappa_{sym}$  are fixed such that $m_{sat}^*/m=0.75$ in symmetric matter and $m_n^*/m-m_p^*/m=-0.1$ in neutron matter.}
\label{tab:epbound}
\end{table*}

The total uncertainty bands in Fig.~\ref{fig:chiralPNM} additionally include the other three sources of uncertainty. The regularization scheme dependence has been explored by explicitly including regulator artifacts for local regulators. Specifically, in Fig.~\ref{fig:chiralPNM}, the neutron-matter uncertainty bands include three different local chiral Hamiltonians which explore short-range $3N$ regulator artifacts; see Ref.~\cite{Lynn:2015jua} for details on the Hamiltonians and Ref.~\cite{Huth:2017wzw} for details on the regulator artifacts. These two sources of uncertainties dominate the total uncertainty band, while the many-body uncertainty is negligible.

To estimate the convergence of the chiral expansion at different densities, the series of expansion coefficients of Eq.~\eqref{eq:chiralExp} can provide insights. In Ref.~\cite{Tews:2018kmu}, we have studied the convergence of the chiral series in pure neutron matter and found it to be reasonable up to a density of $2 n_{\rm sat}$. Beyond that, we expect the chiral expansion to break down even though the expansion parameter only increases by approximately 25\% from $n_{\rm sat}$ to $2 n_{\rm sat}$. Therefore, we restrict the chiral EFT input to densities up to $2 n_{\rm sat}$. In addition, we exclude one chiral Hamiltonian from further consideration because its regulator artifacts lead to a spurious and unphysical attractive $3N$ contribution in neutron matter, as discussed in Ref.~\cite{Tews:2018kmu}. This Hamiltonian represents the lower, soft part of the uncertainty band and  is also in conflict with the unitary-gas bound of Ref.~\cite{Kolomeitsev:2016sjl}, shown in Fig.~\ref{fig:chiralPNM} as a blue dashed line. Excluding this Hamiltonian changes the lower bound of the uncertainty band to the red-dotted line in Fig.~\ref{fig:chiralPNM}, in good agreement with the unitary-gas constraint.

In the following, we use this chiral EFT band up to a density $n_{\text{tr}}$ to constrain two different modelings for the high density equation of state. By varying $n_{\text{tr}}$ from $n_{\text{sat}}$ to $2n_{\text{sat}}$, we will show that, despite the rapid increase of the uncertainty of the neutron-matter EOS with density, chiral EFT constraints remain extremely useful up to $2 n_{\rm sat}$.  

\subsection{The minimal model}
\label{sec:minmod}

The first model that we consider in this analysis, the minimal model or meta-model (MM), assumes the EOS to be smooth enough to be describable in terms of a density expansion about $n_{sat}$. Here, we briefly describe the MM, but see also Refs.~\cite{Margueron:2017eqc,Margueron:2017lup} for more details. 

The MM is described in terms of the empirical parameters of nuclear matter, which are defined as the Taylor coefficients of the density expansion of the energy per particle of symmetric nuclear matter $e_{sat}(n)$ and the symmetry energy $s_{sym}(n)$, 
\begin{align}
e_{sat}(n) &= E_{\text{sat}} + \frac 1 2 K_{\text{sat}} x^2 + \frac 1 6 Q_{\text{sat}} x^3 + \frac 1 {24} Z_{\text{sat}} x^4 + ... \label{eq:esat}\\
s_{sym}(n) &= E_{\text{sym}} + L_{\text{sym}} x+ \frac{1}{2} K_{\text{sym}} x^2 + \frac{1}{6} Q_{\text{sym}} x^3 \nonumber  \label{eq:esym}\\ 
& +\frac{1}{24} Z_{\text{sym}} x^4 + ... \,,
\end{align}
where the expansion parameter $x$ is defined as $x=(n-n_{\text{sat}})/(3n_{\text{sat}})$ and $n=n_n+n_p$ is the baryon density, $n_{n/p}$ are the neutron and proton densities.
A good representation of the energy per particle around $n_{sat}$ and for small isospin asymmetries $\delta=(n_n-n_p)/n$ can be obtained from the following quadratic approximation,
\begin{equation}
e(n,\delta)=e_{sat}(n)+s_{sym}(n)\, \delta^2\, .
\end{equation}
The lowest order empirical parameters can be extracted from nuclear experiments~\cite{Margueron:2017eqc}, but typically carry uncertainties. Especially the symmetry-energy parameters are of great interest to the nuclear physics community and considerable effort is invested into a better estimation of their size.

The MM constructs the energy per nucleon as,
\begin{eqnarray}
e^N(n,\delta)=t^{FG*}(n,\delta)+v^N(n,\delta),
\label{eq:MM:energy}
\end{eqnarray}
where the kinetic energy is expressed as 
\begin{eqnarray}
t^{FG^*}(n,\delta)&=&\frac{t_{sat}^{FG}}{2}\left(\frac{n}{n_{sat}}\right)^{2/3} 
\bigg[ \left( 1+\kappa_{sat}\frac{n}{n_{sat}} \right) f_1(\delta) \nonumber \\
&& \hspace{2.5cm} + \kappa_{sym}\frac{n}{n_{sat}}f_2(\delta)\bigg] ,
\label{eq:MM:kin}
\end{eqnarray}
and the functions $f_1$ and $f_2$ are defined as
\begin{eqnarray}
f_1(\delta) &=& (1+\delta)^{5/3}+(1-\delta)^{5/3} \, , \\
f_2(\delta) &=& \delta \left( (1+\delta)^{5/3}-(1-\delta)^{5/3} \right) .
\end{eqnarray}
The parameters $\kappa_{sat}$ and $\kappa_{sym}$ control the density and asymmetry dependence of the Landau effective mass as ($q$=n or p),
\begin{equation}
\frac{m}{m^*_q(n,\delta)} = 1 + \left( \kappa_{sat} + \tau_3 \kappa_{sym} \delta \right) \frac{n}{n_{sat}} ,
\label{eq:effmass}
\end{equation}
where $\tau_3=1$ for neutrons and -1 for protons.
Taking the limit $\kappa_{sat}=\kappa_{sym}=0$, Eq.~(\ref{eq:MM:kin}) provides the free Fermi gas energy.

The potential energy in Eq.~(\ref{eq:MM:energy}) is expressed as a series expansion in the parameter $x$ and is quadratic in the asymmety parameter $\delta$,
\begin{eqnarray}
v^N(n,\delta)=\sum_{\alpha\geq0}^N \frac{1}{\alpha!}( v_{\alpha}^{sat}+ v_{\alpha}^{sym} \delta^2) x^\alpha u^N_{\alpha}(x) ,
\label{eq:MM:pot}
\end{eqnarray}
where the function $u^N_{\alpha}(x)=1-(-3x)^{N+1-\alpha}\exp(-b n/n_{sat})$ ensures the limit $e^N(n=0,\delta)=0$.
The parameter $b$ is taken large enough for the function $u^N_{\alpha}$ to fall sufficiently fast with density and to not contribute at densities above $n_{sat}$. A typical value is $b=10\ln2\approx 6.93$ such that the exponential function is $1/2$ for $n=n_{sat}/10$.
The MM parameters $v_{\alpha}^{sat}$ and $v_{\alpha}^{sym}$ are simply expressed in terms of the empirical parameters. The MM as expressed in Eqs.(\ref{eq:MM:energy}), (\ref{eq:MM:kin}), and (\ref{eq:MM:pot}) coincides with the meta-model ELFc described in Ref.~\cite{Margueron:2017eqc}, where detailed relations can be found.
To obtain the neutron-star EOS, we extend our models to $\beta$-equilibrium and include a crust as described in Ref.~\cite{Margueron:2017lup}. By varying the empirical parameters within their known or estimated uncertainties, it was shown that the MM can reproduce many existing neutron-star EOS that are based on the assumption that a nuclear description is valid at all densities probed in neutron stars. Therefore, this model is a reliable representation for EOS without exotic phases of matter separated from the nucleonic phase through strong phase transitions.

In the following, the parameter space for the MM will be explored within a Markov-Chain Monte-Carlo algorithm, where the MM parameters are allowed to freely evolve inside the boundaries given in Table.~\ref{tab:epbound}. The resulting models satisfy the chiral EFT predictions in neutron matter for the energy per particle and the pressure up to $n_{\rm tr}$, causality, stability, positiveness of the symmetry energy ($s_{sym}(n)>0$), and also reach the maximum observed neutron-star mass $M_{\rm max}^{\rm obs}$, see the discussion in Sec.~\ref{sec:MMandCSM}. The maximum density associated with each EOS within the MM is given either by the break-down of causality, stability, or positiveness of the symmetry energy condition, or by the end point of the stable neutron-star branch.

\begin{figure*}[t]
\begin{center}
\includegraphics[trim= 0.0cm 0 0 0, clip=,width=0.65\columnwidth]{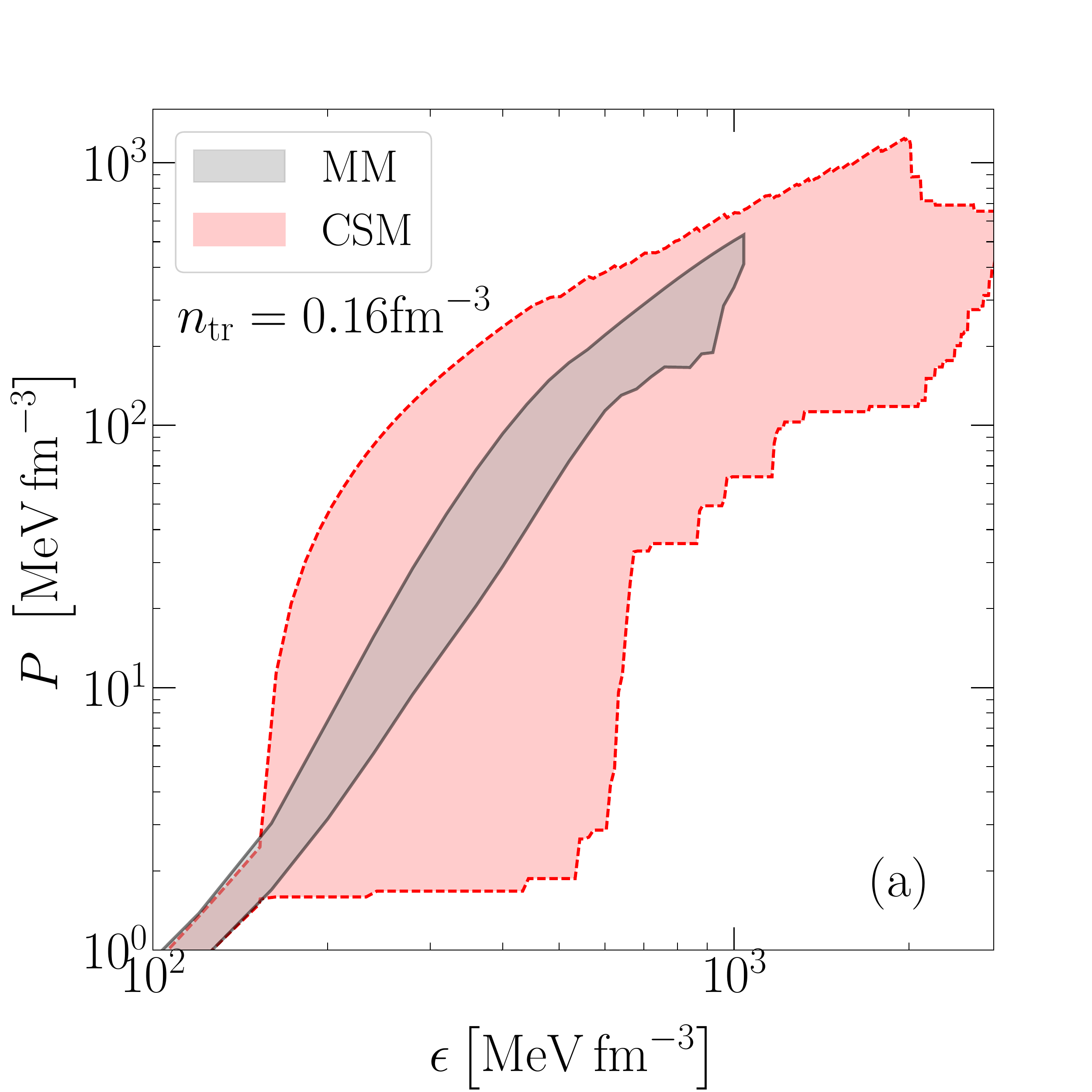}
\includegraphics[trim= 0.0cm 0 0 0, clip=,width=0.65\columnwidth]{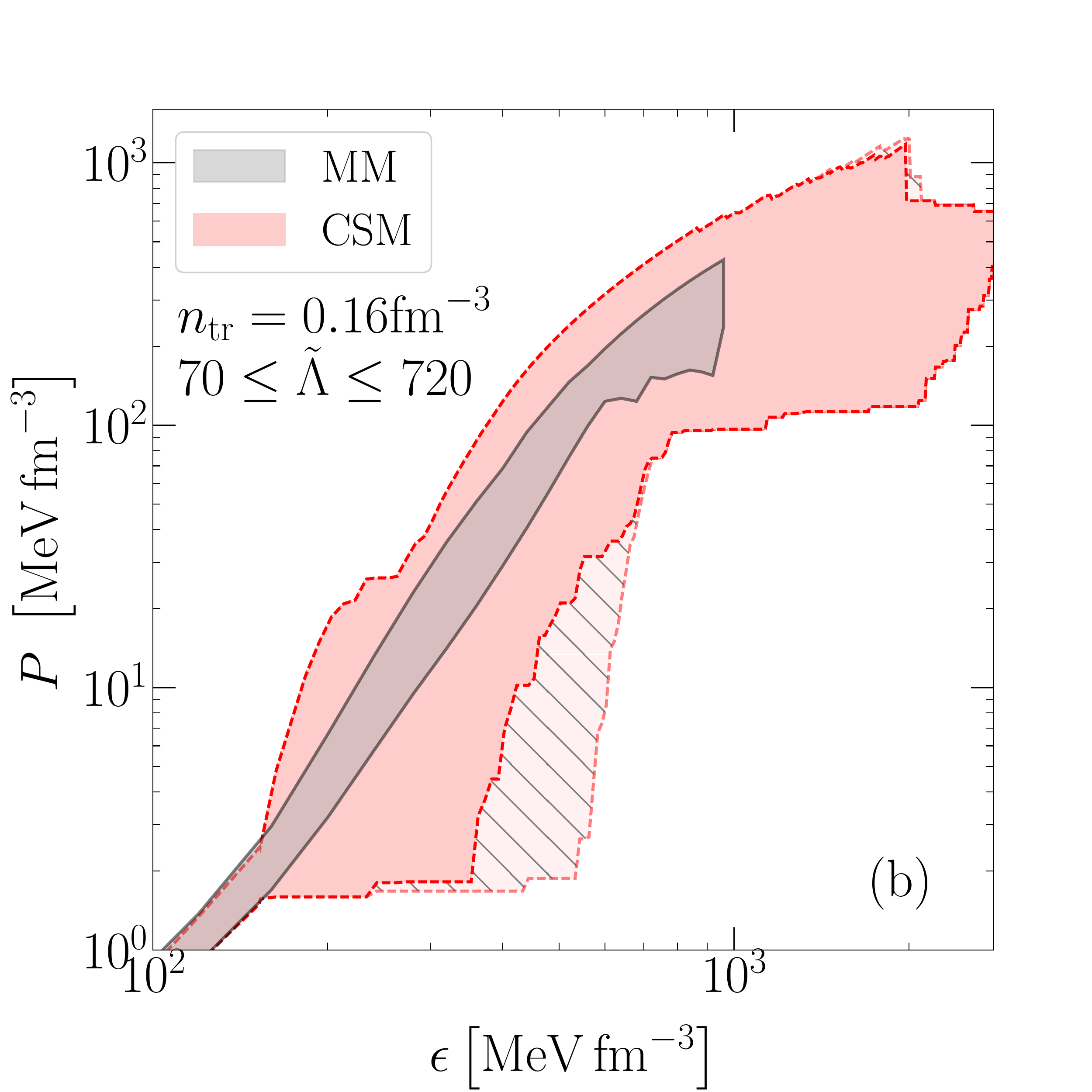}
\includegraphics[trim= 0.0cm 0 0 0, clip=,width=0.65\columnwidth]{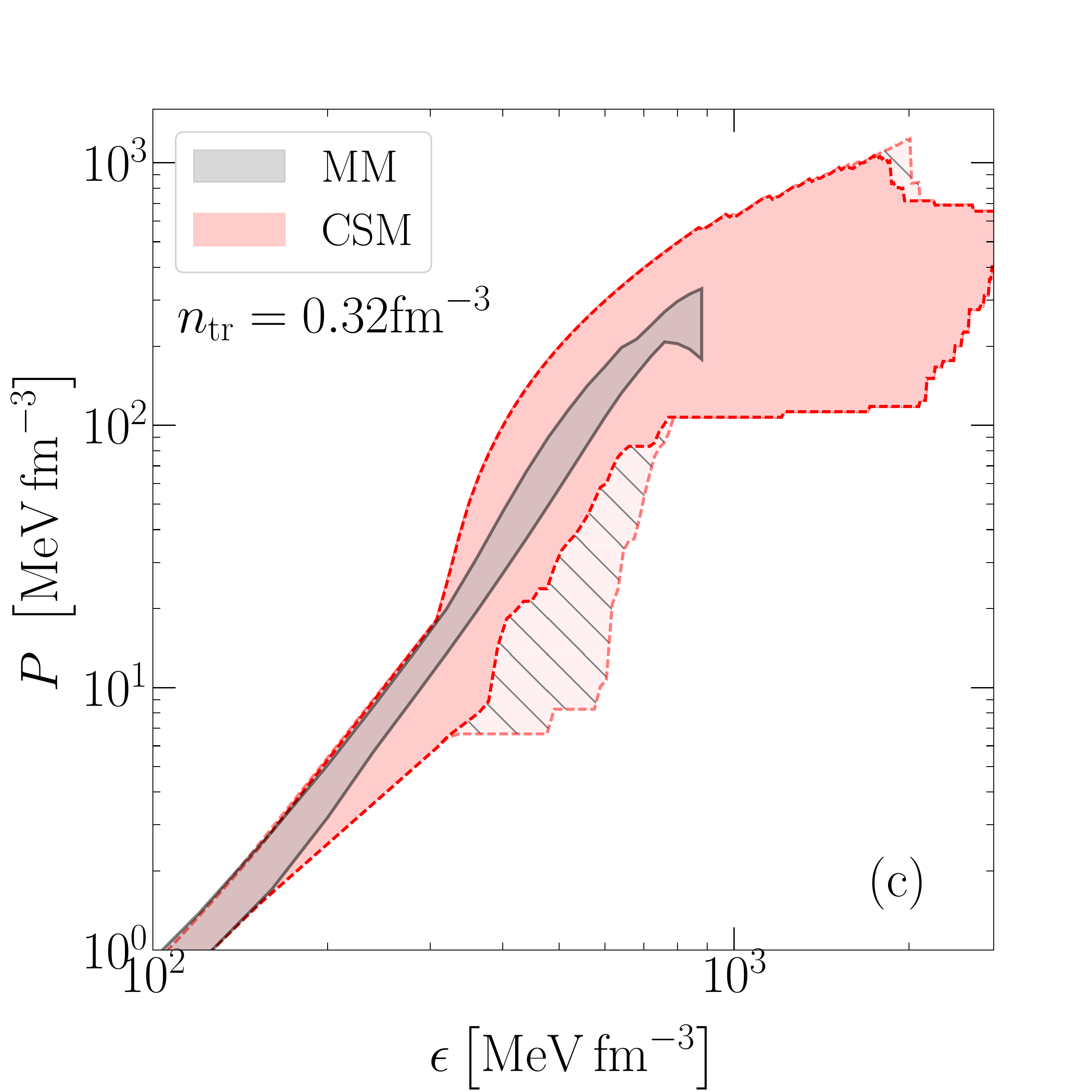}
\caption{\label{fig:EpsPcomp}
Comparison of the allowed EOS envelopes for the MM (black bands) and the CSM (red bands). We show three cases: a) the most general case, where $n_{\text{tr}}=n_{\text{sat}}$ and only $M_{\rm{max}}\geq 1.9 M_{\odot}$ is enforced, b) for $n_{\text{tr}}=n_{\text{sat}}$ when enforcing $70\leq \tilde{\Lambda} \leq 720$ and c) for $n_{\text{tr}}=2 n_{\text{sat}}$. When additionally enforcing $R_{1.6}\geq 10.68$ km~\cite{Bauswein:2017vtn}, the hatched regions are excluded.
}
\end{center}
\end{figure*}

\subsection{The maximal model}
\label{sec:maxmod}

The second model that we consider in this analysis, the maximal model (CSM), is based on an extension of the speed of sound in neutron-star matter. Starting from the pure neutron matter calculations, we construct the neutron-star EOS up to $n_{\rm tr}$ by constructing a crust as described in Ref.~\cite{Tews:2016ofv} and extending the neutron-matter results to $\beta$ equilibrium above the crust-core transition. Having constructed the EOS up to $n_{\rm tr}$ we compute the speed of sound, 
\begin{equation}
c_S^2 = \frac{\partial p(\epsilon)}{\partial \epsilon}\,,
\end{equation}
where $p$ is the pressure and $\epsilon$ is the energy density. Above $n_{\rm tr}$, we parametrize the speed of sound in a very general way: we randomly sample a set of points $c_S^2(n)$, where the values for $c_S$ have to be positive and are limited by the speed of light (stability and causality), and interpolate between the different sampling points using linear segments. The individual points are randomly distributed in the interval $n_{\rm tr}-12 n_{\rm sat}$.  From the resulting speed-of-sound curve, we reconstruct the EOS step-by-step starting at $n_{\text{tr}}$, where $\epsilon(n_{\text{tr}})$,  
$p(n_{\text{tr}})$, and $\epsilon'(n_{\text{tr}})$ are known:
\begin{align}
n_{i+1}&= n_i + \Delta n \\
\epsilon_{i+1} &= \epsilon_i +\Delta\epsilon= \epsilon_i + \Delta n \cdot \left(\frac{\epsilon_i+p_i}{n_i}\right) \\
p_{i+1} &= p_i + c_S^2 (n_i) \cdot \Delta \epsilon\,,
\end{align}
where $i=0$ defines the transition density $n_{\text{tr}}$. In the second line we have used the thermodynamic relation $p=n \partial \epsilon/\partial n -\epsilon$, which is valid at zero temperature. 

In that way, we iteratively obtain the high-density EOS. We have explored extensions for a varying number of $c_S^2(n)$ points, i.e., for 5-10 points, and found that the differences between these extensions are marginal. We, therefore, choose 6 sampling points. For each sampled EOS, we generate a second version which includes a strong first-order phase transition with a random onset density and width, to explicitly explore such extreme density behavior.

The CSM for neutron-star applications was introduced in Ref.~\cite{Tews:2018kmu}, and represents and extension of the model of Ref.~\cite{Alford:2013aca}. A similar model was used in Ref.~\cite{Greif:2018njt}. However, in contrast to Ref.~\cite{Tews:2018kmu} we have extended this model to explore the complete allowed parameter space for the speed of sound, by abandoning the specific functional form of Ref.~\cite{Tews:2018kmu} in favor of an extension using linear segments. This more conservative choice leads to slightly larger uncertainty bands, but allows us to make more definitive statements about neutron-star properties. The resulting EOS parameterizations represent possible neutron-star EOS and may include drastic density dependences, e.g., strong phase transitions which lead to intervals with a drastic softening or stiffening of the EOS. 
This represents a stark contrast to the MM, which does not include such behavior, and might give insights into the constituents of neutron-star matter at high-densities. The predictions of the CSM represent the widest possible domain for the respective neutron-star observables consistent with the low density input from chiral EFT. If observations outside of this domain were to be made, this would imply a breakdown of nuclear EFTs at densities below the corresponding $n_{\rm tr}$. 

Since the CSM represents very general EOSs only governed by the density dependence of the speed-of-sound, it does not allow any statements about possible degrees of freedom. In this sense, it is very similar to extensions using piecewise polytropes which were introduced in Ref.~\cite{Read:2008iy} and have been used extensively to determine neutron-star properties; see, e.g., Ref.~\cite{Hebeler:2013nza,Raithel:2016bux,Annala:2017llu}. However, in contrast to polytropic extensions, in the CSM the speed of sound is continuous except when first-order phase transition are explicitly accounted for. Discontinuities in the speed of sound affect the study of tidal polarizabilities, where $c_S^{-1}$ enters, by introducing features whose source is solely the choice of parametrization.

\subsection{Comparison of MM and CSM}
\label{sec:MMandCSM}

\begin{figure*}[t]
\begin{center}
\includegraphics[trim= 0.0cm 0 0 0, clip=,width=0.65\columnwidth]{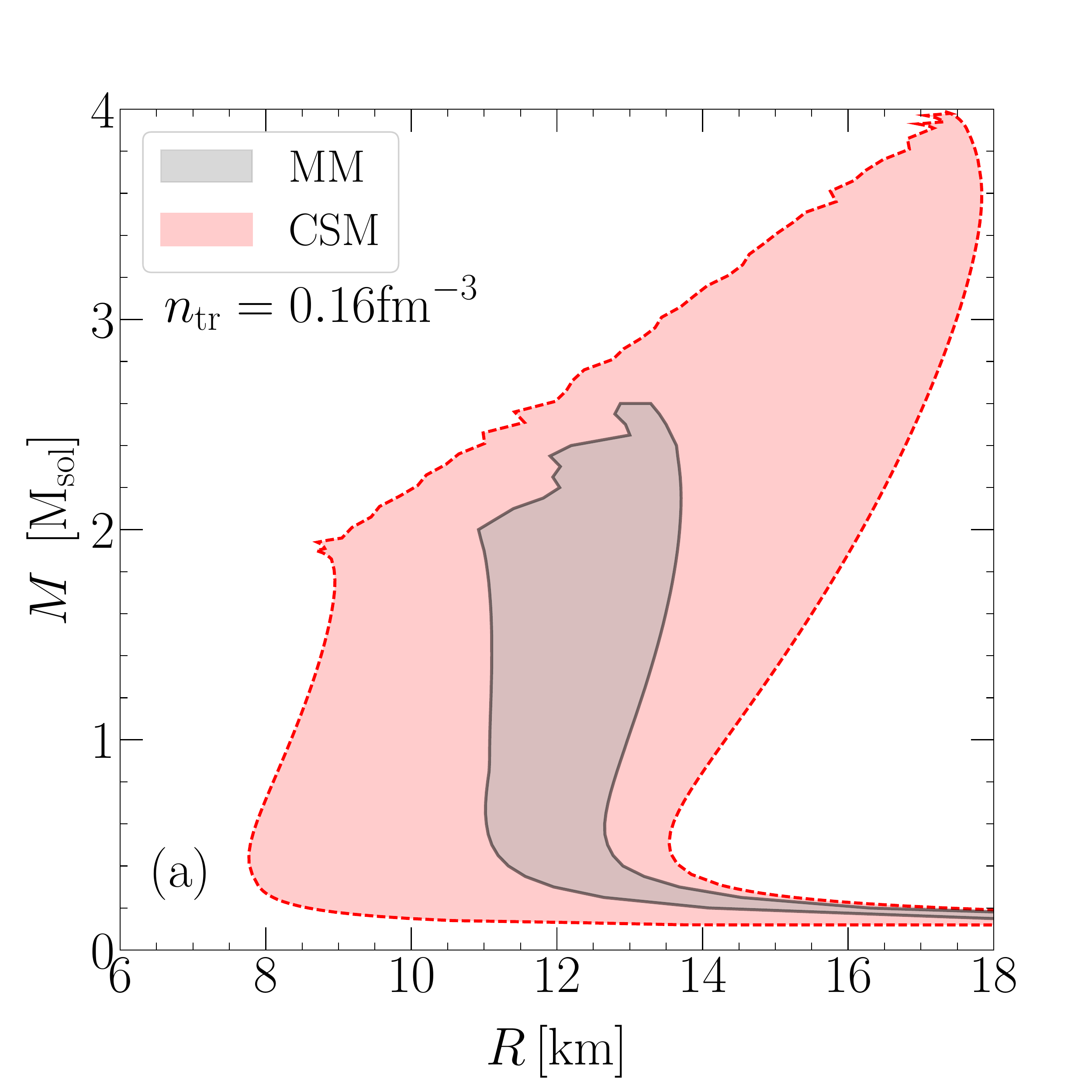}
\includegraphics[trim= 0.0cm 0 0 0, clip=,width=0.65\columnwidth]{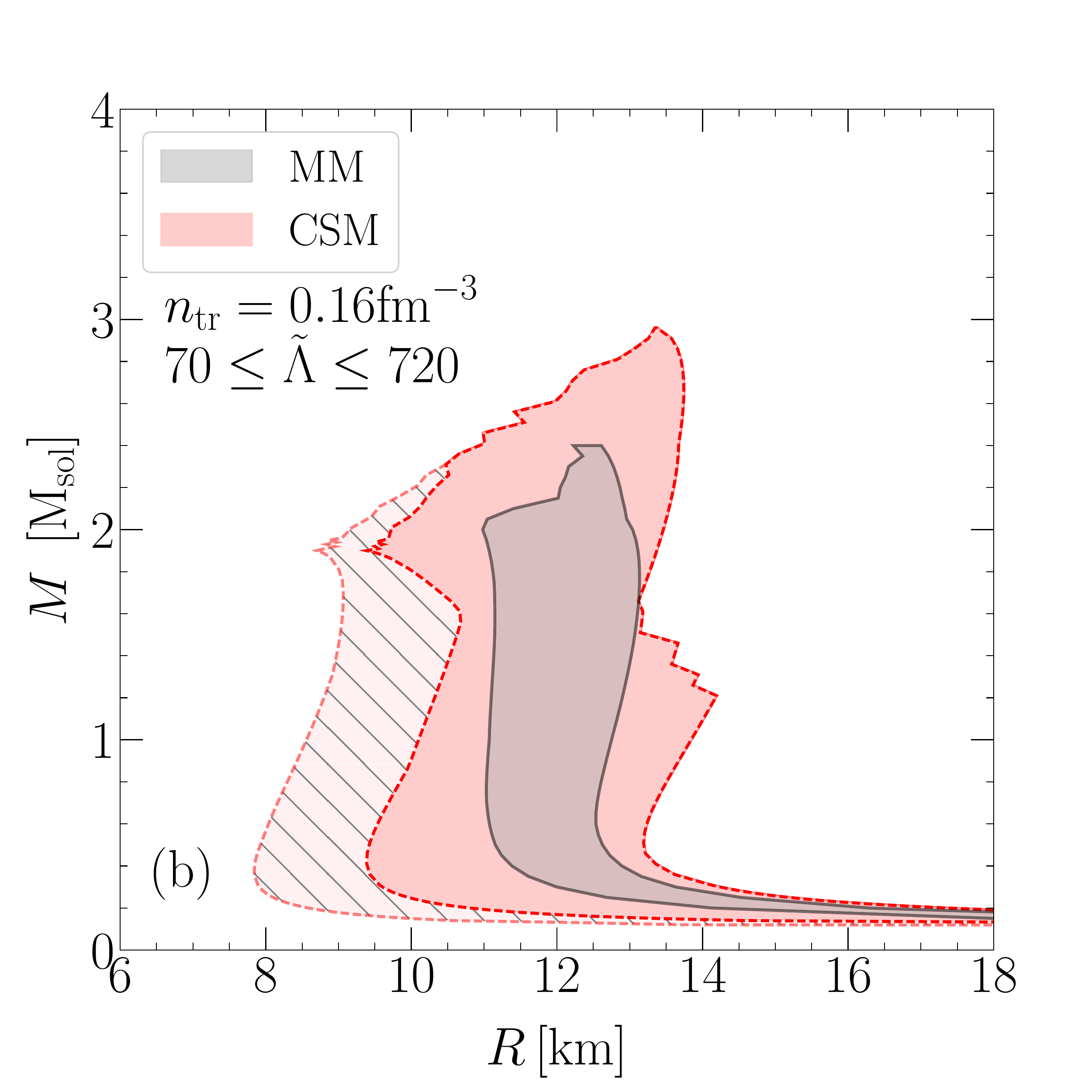}
\includegraphics[trim= 0.0cm 0 0 0, clip=,width=0.65\columnwidth]{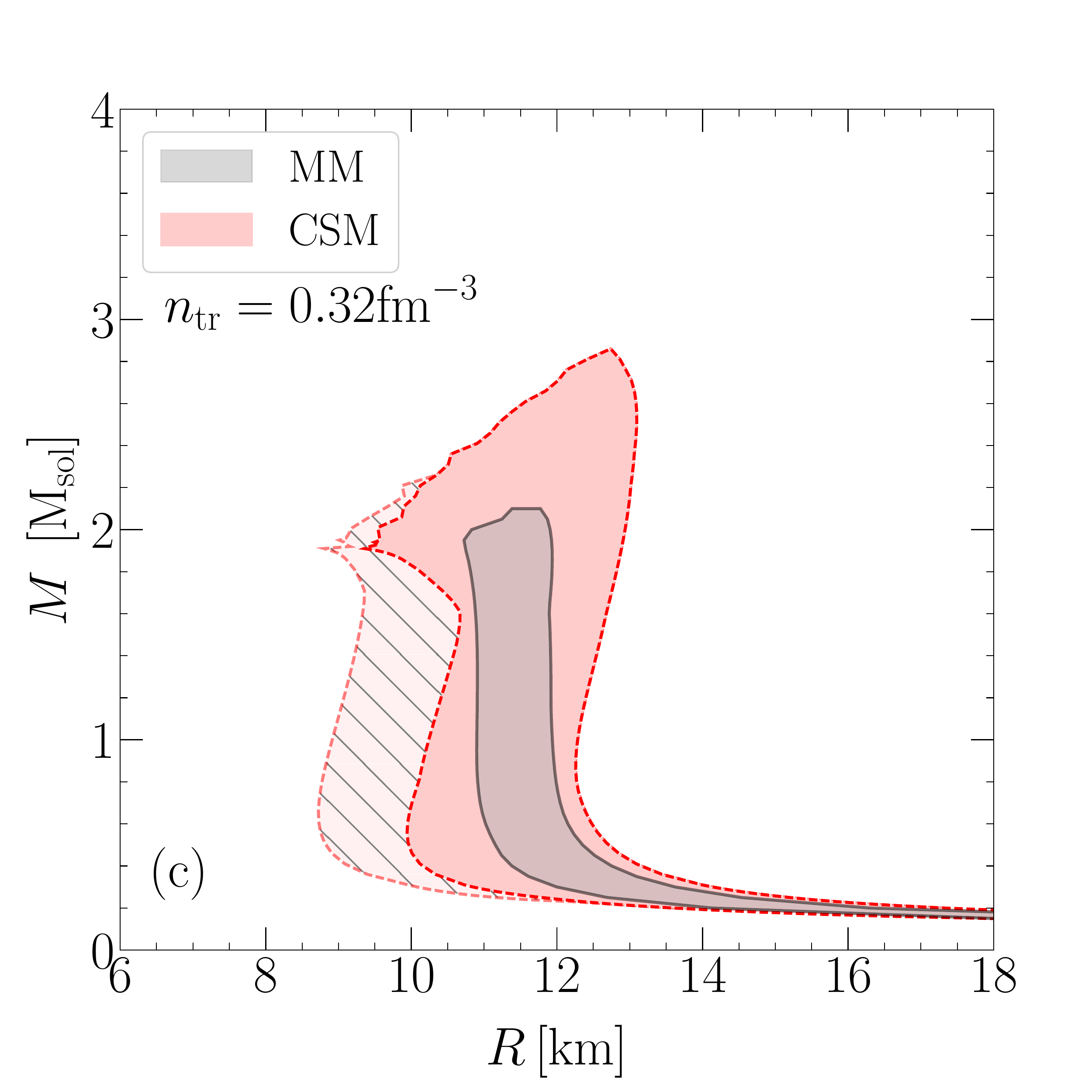}
\caption{\label{fig:MRcomp}
Comparison of the allowed MR envelopes for the MM (black bands) and the CSM (red bands). We show three cases: a) the most general case, where $n_{\text{tr}}=n_{\text{sat}}$ and only $M_{\rm{max}}\geq 1.9 M_{\odot}$ is enforced, b) for $n_{\text{tr}}=n_{\text{sat}}$ when enforcing $70\leq \tilde{\Lambda} \leq 720$, and c) for $n_{\text{tr}}=2 n_{\text{sat}}$. When additionally enforcing $R_{1.6}\geq 10.68$ km~\cite{Bauswein:2017vtn}, the hatched regions are excluded.
}
\end{center}
\end{figure*} 

For both the MM and CSM we generate thousands of EOSs that are consistent with low-density constraints from chiral EFT. In addition, the observations of heavy two-solar-mass pulsars in recent years~\cite{Demorest2010,Antoniadis2013,Fonseca2016} place important additional constraints on these EOSs, which we enforce by requiring  $M_{\text{max}}>M_{\rm max}^{\rm obs}$ for all our EOSs. 
To be conservative, as the limit for $M_{\rm max}^{\rm obs}$ we choose the centroid of the maximum observed mass minus twice the error-bar on the observation. For the two heaviest neutron stars observed up to now~\cite{Demorest2010,Antoniadis2013,Fonseca2016}, this gives $M_{\rm max}^{\rm obs}\approx 1.9 M_\odot$. 

We now compare the predictions of both the MM (black bands with solid contour) and CSM (red bands with dotted contour) for the EOS of neutron-star matter, see Fig.~\ref{fig:EpsPcomp}, and the mass-radius (MR) relation, see Fig.~\ref{fig:MRcomp}. In the respective figures, we show the EOS and MR envelopes for $n_{\rm tr}=n_{\rm sat}$ [panels (a)] and for $n_{\rm tr}=2 n_{\rm sat}$ [panels (c)], where ragged edges are due to the limited number of models. In all cases, the MM is a subset of the CSM, as expected. Also, the two models, which treat the neutron-star crust with different prescriptions, show excellent agreement at low densities. For $n_{\rm tr}=n_{\rm sat}$, the MM and CSM EOSs agree very well up to  $n_{\rm tr}$, while for $n_{\rm tr}=2 n_{\rm sat}$ the MM only samples a subset of the chiral EFT input, because the $M_{\rm max}^{\rm obs}$ constraint forces the EOS to be sufficiently stiff which excludes the softest low-density neutron-matter EOS. This is a consequence of the smooth density expansion around $n_{\rm sat}$ in the MM. In the CSM, instead, a non-smooth stiffening of these softest EOS at higher densities can help stabilize heavy neutron stars, which is why the complete low-density band from chiral EFT is sampled.  We also find that going from $n_{\rm tr}=n_{\rm sat}$ to $n_{\rm tr}=2 n_{\rm sat}$ allows to considerable reduce the EOS uncertainty for the CSM. The MM uncertainty is also slightly reduced and the MM band gets narrower. These results show that even though the theoretical uncertainties in the neutron-matter EOS increase fast in the density range $1-2 n_{\text{sat}}$, the additional information provided allows to substantially reduce uncertainties in the CSM EOS:   essentially, the chiral EFT constraint excludes the possibility of phase transitions in the region going from 1 to $2n_{sat}$. The impact of phase transitions above $2n_{sat}$ on the EOS is very much reduced compared to the case where they are allowed to appear at lower densities, because we impose the $M_{\rm max}^{\rm obs}$ constraint. A large domain of soft CSM EOSs is, thus, excluded. The stiff MM and CSM EOS are very close up to $2n_{sat}$, as expected.

These observations are also reflected in the MR predictions of both models. For $n_{\rm tr}=n_{\rm sat}$ [panel (a)], the CSM (MM) leads to a radius range of a typical neutron star of $1.4 M_{\odot}$ of $8.4-15.2$ km ($10.9-13.5$ km). This range reduces dramatically for $n_{\rm tr}=2 n_{\rm sat}$ [panel (c)], where we find $8.7-12.6$ km ($10.9-12.0$ km) for the CSM (MM). 

In the last case, the radius uncertainty for a typical neutron star is only about 1 km in the MM, compatible with the expected uncertainty of the NICER mission~\cite{NICER1}. This allows for a possible tight confrontation between the MM and the NICER results. If such an observation should be made in the near future, we will be able to better constrain dense-matter phase transitions. In contrast, the CSM, which includes EOS with sudden softening or stiffening at higher densities, dramatically extends the allowed envelopes for the EOS and the MR relation as compared with the MM. These differences in the predictions of the MM and CSM can be used to identify regions for the neutron-star observables, for which statements about the constituents of matter might be possible. For example, the observation of a typical neutron star with a radius of 10 km would imply the existence of a softening phase transition, that would hint on new phases of matter appearing in the core of neutron stars. Instead, in regions were both the MM and CSM agree, the masquerade problem does not allow statements about the constituents of neutron-star matter at high densities~\cite{Alford:2004pf}.

In Fig.~\ref{fig:MRcomp}, the maximum mass for $n_\mathrm{tr}=n_\mathrm{sat}$ is almost $4M_\odot$ while it is only $2.9M_\odot$ if $n_\mathrm{tr}=2n_\mathrm{sat}$.
It is interesting to compare these findings with previous predictions for the maximum mass of neutron stars. Connecting a nucleonic EOS to the stiffest possible EOS at $n_\mathrm{tr}=2n_\mathrm{sat}$, the maximum mass was predicted to be  $2.9 M_\odot$~\cite{Rhoades1974}, as in our case. With a similar approach but defining $n_\mathrm{tr}$ to lie between 1 and $2n_\mathrm{sat}$, Ref.~\cite{Kalogera1996} predicted the maximum mass to be  $3.2 M_\odot$. Note, however, that by lowering $n_\mathrm{tr}$, the authors found $3.9 M_\odot$ as the maximum mass, again very close to our prediction. The maximum mass of neutron stars is therefore tightly correlated with $n_\mathrm{tr}$ for both the MM and CSM models, as shown in Fig.~\ref{fig:MRcomp}.

Finally, due to the rather soft density dependence of chiral EFT constraints in the density range $1-2 n_{\rm sat}$, $n_{\rm tr}=2 n_{\rm sat}$ together with the constraint $M_{\text{max}}>M_{\rm max}^{\rm obs}$ seems to strongly disfavor EOS that lead to the appearance of disconnected compact-star branches, as suggested in Ref.~\cite{Paschalidis:2017qmb}. Such EOS need very strong first-order phase transitions, which would soften the EOS so much that heavy two-solar-mass neutron stars cannot be supported, in accordance  with the findings in Ref.~\cite{Alford:2015dpa}. Instead, chiral EFT calculations up to $n_{\rm tr}=2 n_{\rm sat}$ imply that EOSs with first-order phase transitions lead to neutron stars of the classification "A" or "C" of Ref.~\cite{Alford:2013aca}. 

\section{Results for GW170817}\label{sec:results}

In this section, we confront the recent neutron-star merger observation GW170817 by the LIGO-Virgo (LV) collaboration with our two classes of EOS models. 

\subsection{Posterior of the LIGO-Virgo analysis}
\label{sec:posterior}

The LV collaboration observed the GW signal of GW170817 for about $100 s$ (several 1000 revolutions, starting from 25 Hz) and performed detailed analyses of the wave front~\cite{Abbott:2018wiz}. Because the chirp mass $M_{\text{chirp}}$, defined as 
\begin{equation}
M_{\text{chirp}}=\frac{(m_1 m_2)^{3/5}}{(m_1+m_2)^{1/5}}\,,
\end{equation} 
can be extracted from the entire signal, this observation allowed to put tight constraints on it. For GW170817, the LV collaboration precisely determined $M_{\text{chirp}}= 1.186\pm 0.001 M_{\odot}$.

\begin{figure}[t]
\includegraphics[trim= 0.0cm 0 0 0, clip=,width=0.9\columnwidth]{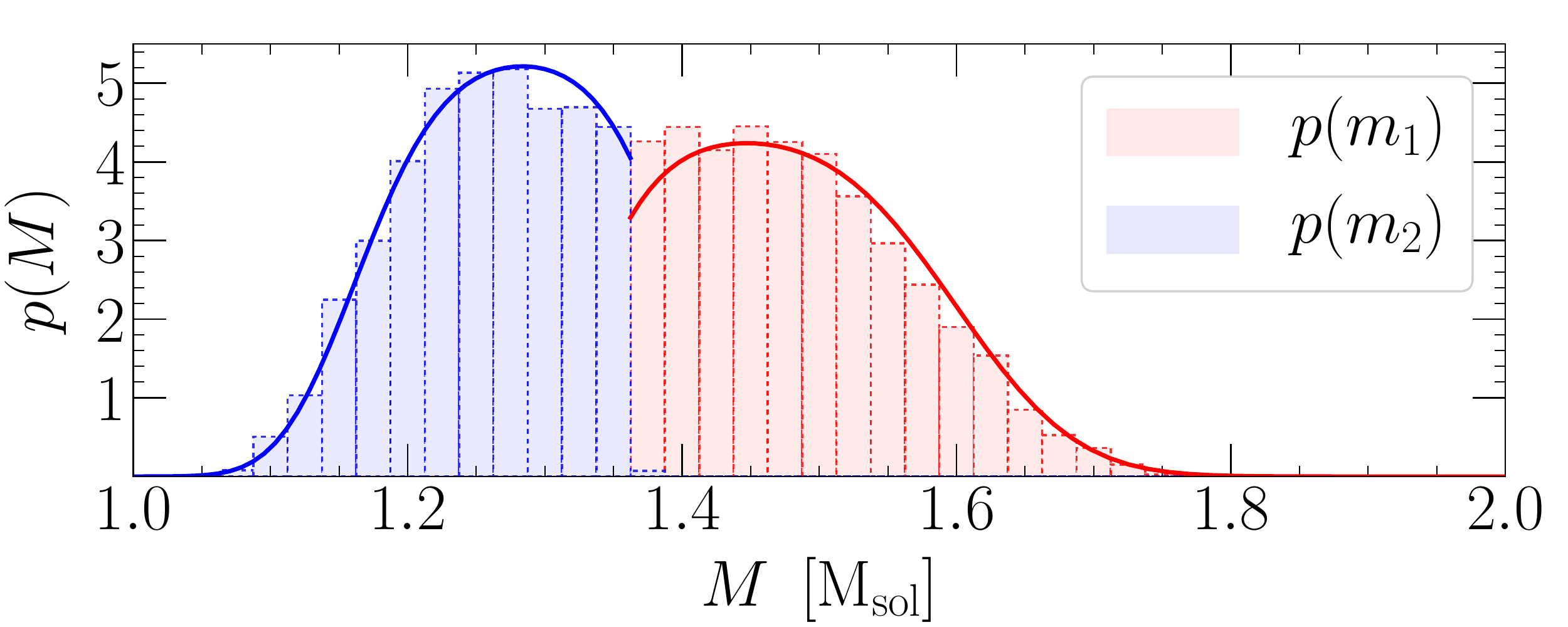}\\
\includegraphics[trim= 0.0cm 0 0 0, clip=,width=0.9\columnwidth]{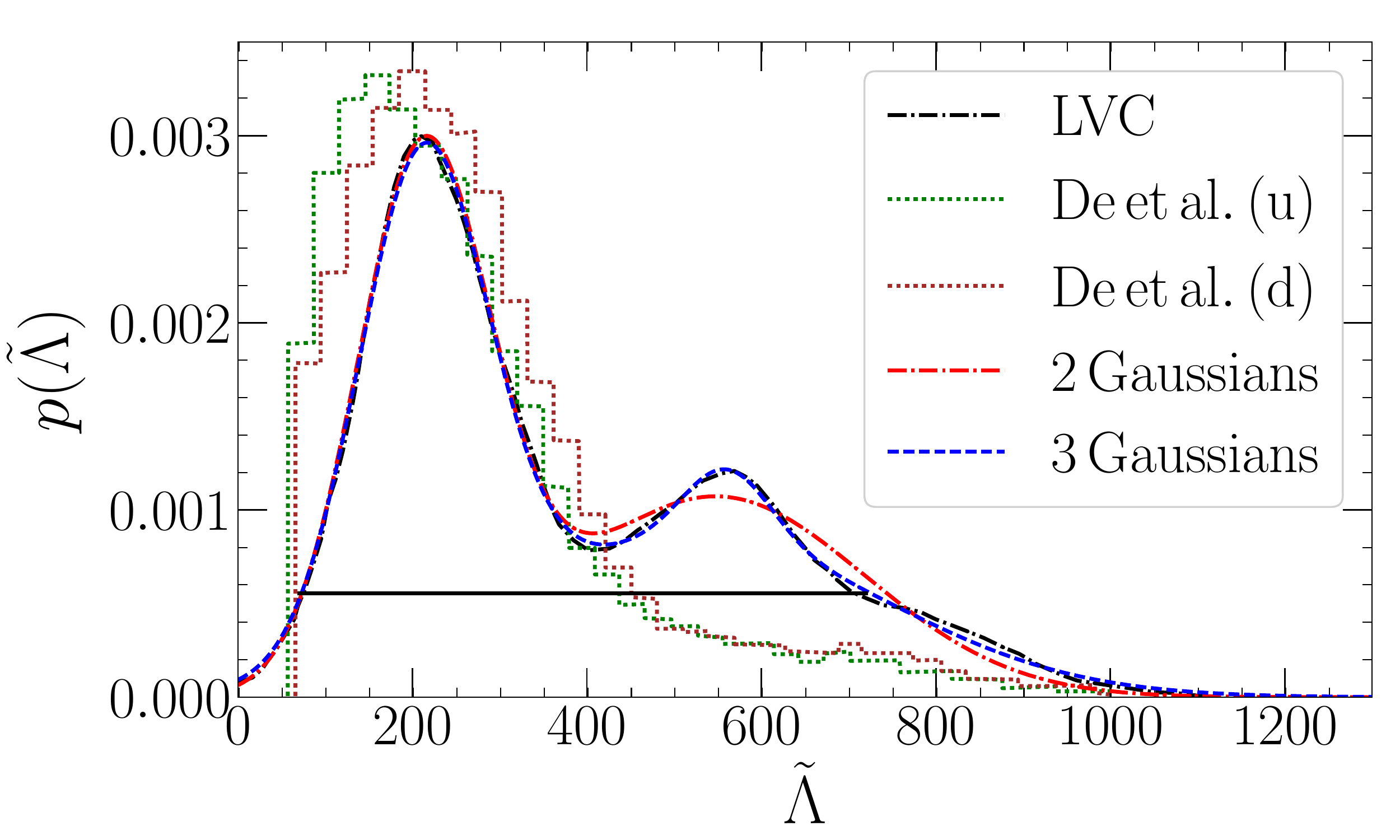}
\caption{\label{fig:posteriors}
Posteriors for the LV observation of GW170817. Upper panel: The mass distributions for $m_1$ and $m_2$ from Ref.~\cite{Abbott:2018wiz} (histograms) and the distributions used in this work (solid lines), see Eq.~\eqref{eq:massdist}. Lower panel: Marginalized and normalized posterior probability for the distribution $p(\tilde{\Lambda})$ as defined in this work. We also show the corresponding distributions for the analysis of the LV collaboration (LVC), and the reanalysis of Ref.~\cite{De:2018uhw} for the two extreme cases [uniform mass prior (u) and mass prior informed by double neutron stars (d)].
}
\end{figure}

\begin{table*}
\caption{\label{tab:LVCLtilde}
Fit parameters of the Gaussians of Eq.~\eqref{eq:Gaussians}}
\centering
\begin{tabularx}{\textwidth}{XXXXXXXXXX}
\hline
\hline
N & $a_1$ & $\Lambda_1$ & $\sigma_1$ & $a_2$ &$\Lambda_2$ &$\sigma_2$ & $a_3$ &$\Lambda_3$ & $\sigma_3$\\
\hline
2  & 281.6 & 212.6 & 76.2 & 106.5
& 547.5 & 171.0 &  & &\\
3 & 266.6 & 212.4 & 74.2 & 85.0 & 523.6 & 219.2 & 38.6
& 560.8 & 49.5\\
\hline
\hline
\end{tabularx}
\end{table*}

The extraction of higher-order GW parameters from the wavefront is complicated for several reasons. First, higher-order parameters are sensitive to the GW signal at later times and, thus, only a smaller part of the signal is suitable for their extraction. Second, there exist ambiguities between different higher-order parameters, e.g., between the individual neutron-star spins and the tidal polarizability. Because of this, the LV collaboration provided results for both a low-spin and a high-spin scenario. In this work, we only investigate the low-spin scenario for two reasons. First, large spins are not expected from the observed galactic binary NS population. Second, because neutron stars spin down over time, low spins are also expected from the extremely long merger time of GW170817 of the order of gigayears. Therefore, the low spin scenario is expected to be the more realistic scenario for binary neutron-star mergers such as GW170817.

The above mentioned problems in the extraction of higher-order parameters lead to weaker constraints on the individual masses of the two component neutron stars in GW170817. With $m_1$ being the mass of the heavier and $m_2$ being the mass of the lighter neutron star in the binary, the mass distribution of the individual stars is typically described in terms of the parameter $q=m_2/m_1$. The observed mass distributions for $m_1$ and $m_2$ are presented as histograms in the upper panel of Fig.~\ref{fig:posteriors}. To use this information in our calculations, we describe the posterior of the LV collaboration for $M_{\text{chirp}}$ and $q$ by the analytical probability distribution~\cite{Margalit:2017}
\begin{equation}
p(q,M_{\text{chirp}}) = p(q) p(M_{\text{chirp}})\,,
\end{equation}
where
\begin{equation}
p(M_{\text{chirp}}) \propto \exp [- (M_{\text{chirp}}-\bar{M}_{\text{chirp}})^2/2\sigma_{M}^2]\,,
\end{equation}
with $\bar{M}_{\text{chirp}}=1.186M_{\odot}$ and $\sigma_{M}= 10^{-3}M_{\odot}$~\cite{Abbott:2018wiz}. For the mass asymmetry $q$, we have fitted the function 
\begin{equation}
p(q)=\exp \left(-\frac12 v(q)^2 -\frac{c}{2} v(q)^4 \right)\,,\label{eq:massdist}
\end{equation}
to the LV posterior for the component masses. We find $c=1.83$ and $v(q)=(q-0.89)/0.20$, and compare the resulting normalized analytic distributions with the observed data in the upper panel of Fig.~\ref{fig:posteriors}.

Since in this work we will confront the gravitational-wave observations of the LV collaboration with nuclear physics constraints, i.e., use our set of EOSs together with the source properties of GW170817 to postdict the distribution of $\tilde{\Lambda}$, we do not make use of the observed probability distribution for $\tilde{\Lambda}$. However, for reasons of completeness, we have fitted functions consisting of two and three Gaussians of the form
\begin{eqnarray}
p(\tilde{\Lambda}) = \sum_{i=1}^N a_i e^{-\frac 1 2 \left( \frac{\tilde{\Lambda}-\Lambda_i}{\sigma_{i}}\right)^2}\label{eq:Gaussians}
\end{eqnarray}
to the observed LV posterior for $\tilde{\Lambda}$. The resulting parameters $a_i$, $q_i$ and $\sigma_{qi}$ are reported in Table~\ref{tab:LVCLtilde}, and the resulting functions as well as the LV result are plotted in the lower panel of Fig.~\ref{fig:posteriors}, where the horizontal black line represents the 90\% LV confidence level for $\tilde{\Lambda}$. We also show the posteriors for the reanalysis of Ref.~\cite{De:2018uhw} for the two extreme cases [uniform mass prior (u) and mass prior informed by double neutron stars (d)]. The main difference between the two analyses lies in the appearance of a second peak in the posterior probability distribution around $\tilde{\Lambda}\sim 600$ for the LV result. The origin of this second peak is not well understood: the peak may be washed out considering a wider domain of frequencies, starting from 23~Hz as in Ref.~\cite{De:2018uhw}. The presence of the second peak is indeed an important issue for the prediction of $\tilde{\Lambda}$: including the second peak, the upper boundary for the 90\%-CL is 720, while it drops if the second peak is absent.

Therefore, in the following, we consider a structureless flat probability distribution in $\tilde{\Lambda}$, and sample the mass distributions for $m_1$ and $m_2$ in GW170817 from the analytic function $p(q,M_{\text{chirp}})$.

\subsection{Areas of constant $\Lambda$}

\begin{figure}[t]
\includegraphics[trim= 0.0cm 0 0 0, clip=,width=0.9\columnwidth]{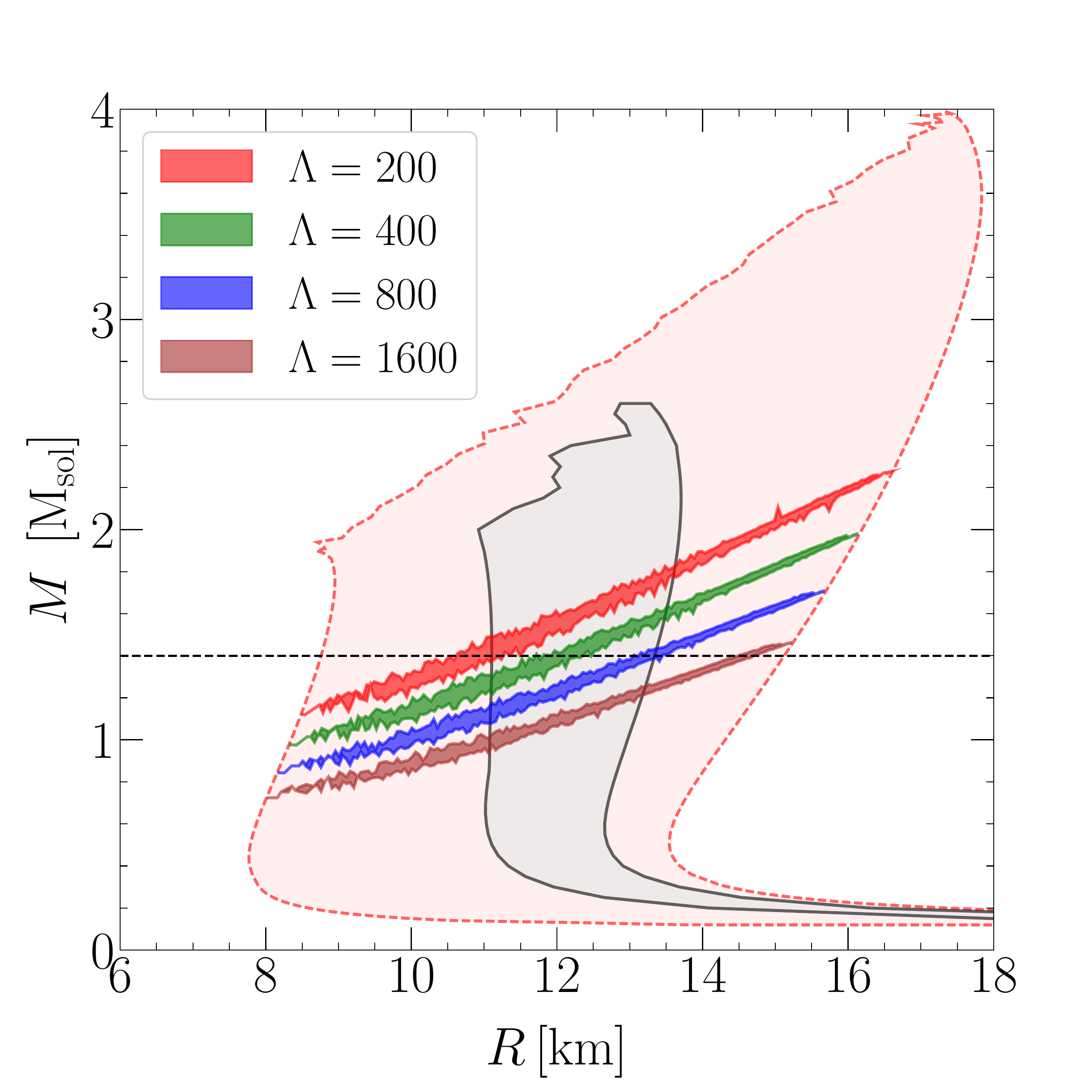}
\caption{\label{fig:MRLam}
Mass-radius envelopes for $n_{\rm tr}=n_{\rm sat}$ of Fig.~\ref{fig:MRcomp}(a) and areas of constant $\Lambda$ for all CSM EOS parametrizations. We show areas for $\Lambda=200$ (red), $\Lambda=400$ (green), $\Lambda=800$ (blue), and for $\Lambda=1600$ (brown). For a typical $1.4 M_{\odot}$ neutron star (horizontal dashed line), a constraint on $\Lambda$ is equivalent to a radius constraint. The corresponding values for the MM (not shown) always lie withing the areas for the CSM. 
}
\end{figure} 

Before addressing GW170817, we focus on the tidal polarizability $\Lambda$ of individual neutron stars. The tidal polarizability describes how a neutron star deforms under an external gravitational field, and depends on neutron-star properties as
\begin{align}
\Lambda &=\frac23 k_2 \left(\frac{c^2}{G} \frac{R}{M}\right)^5\,.
\end{align}
Here, $k_2$ is the tidal love number, that is computed together with the Tolman-Oppenheimer-Volkoff equations; see, for example, Refs.~\cite{Flanagan2008,Damour2009,Moustakidis:2016sab} for more details.

\begin{figure*}[t]
\includegraphics[trim= 0.0cm 0 0 0, clip=,width=0.67\columnwidth]{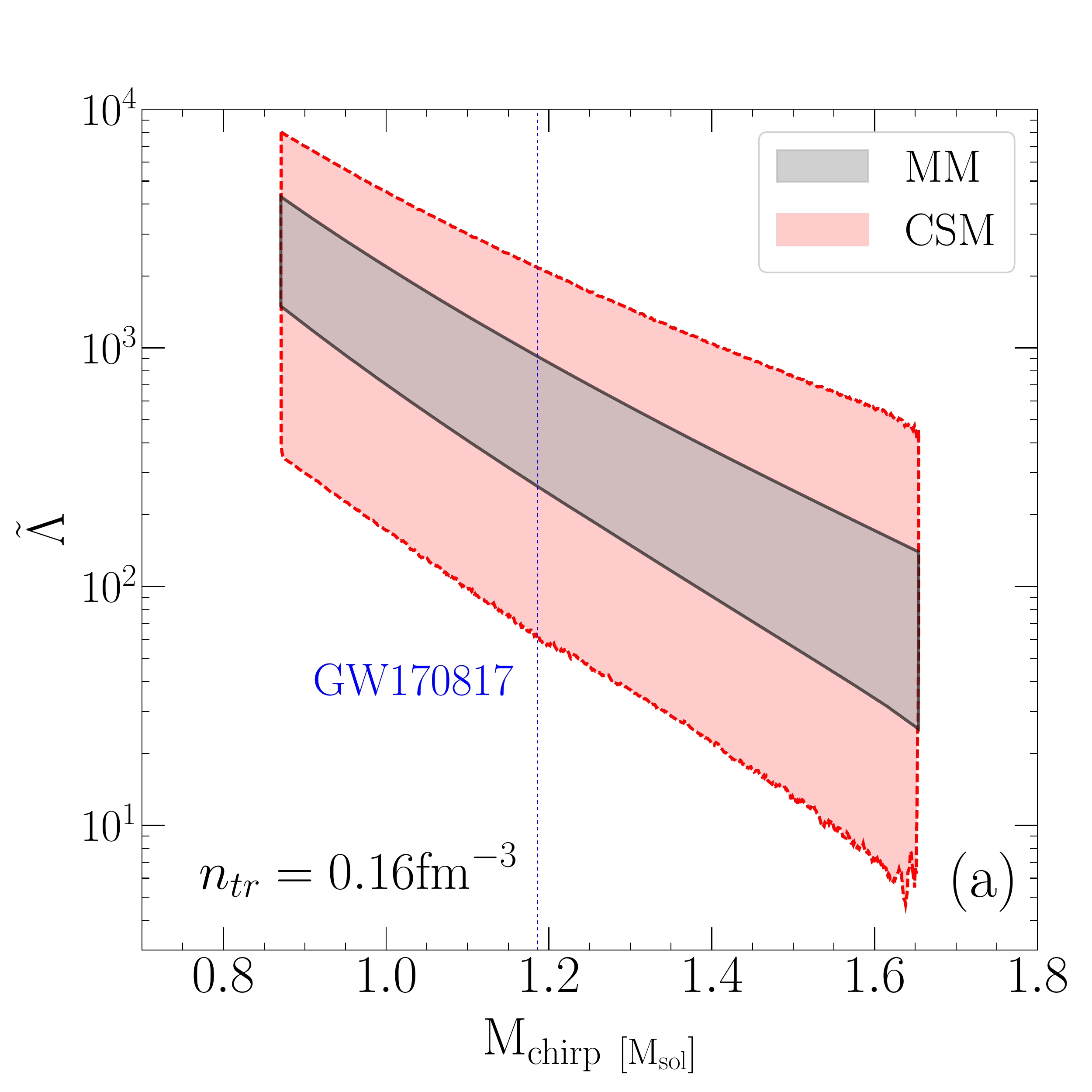}
\includegraphics[trim= 0.0cm 0 0 0, clip=,width=0.67\columnwidth]{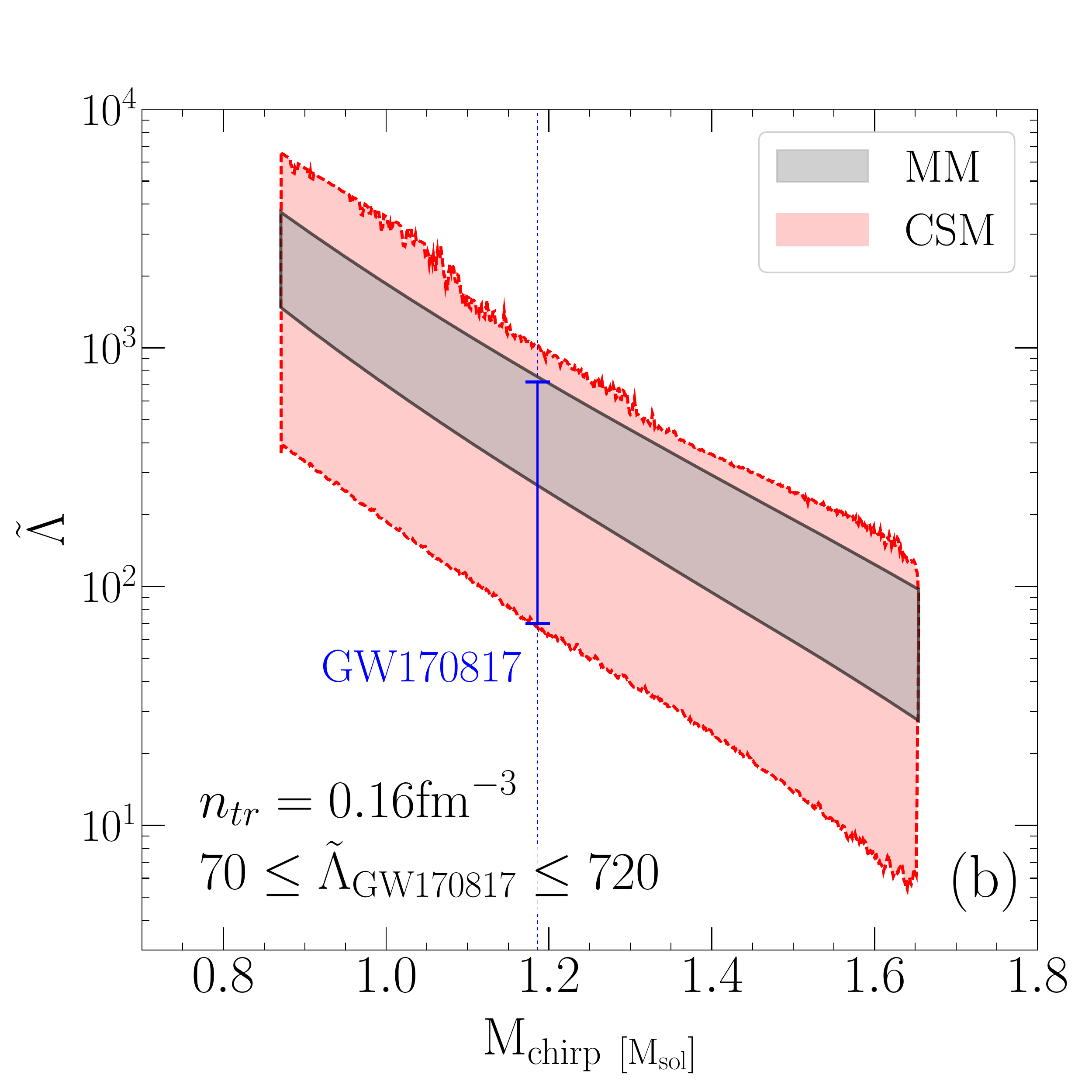}
\includegraphics[trim= 0.0cm 0 0 0, clip=,width=0.67\columnwidth]{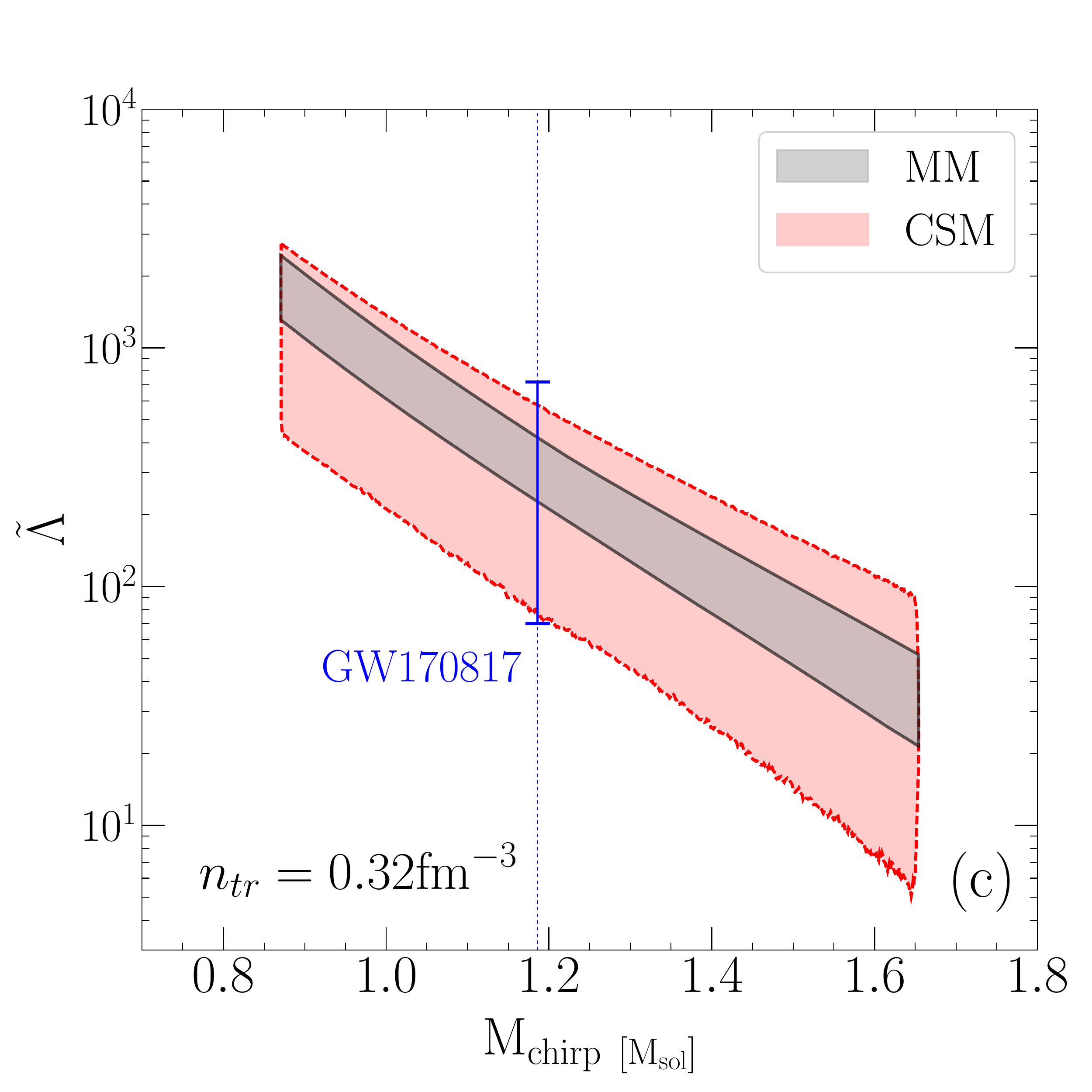}\\
\null\hfill
\includegraphics[trim= 0.0cm 0 0 0, clip=,width=0.67\columnwidth]{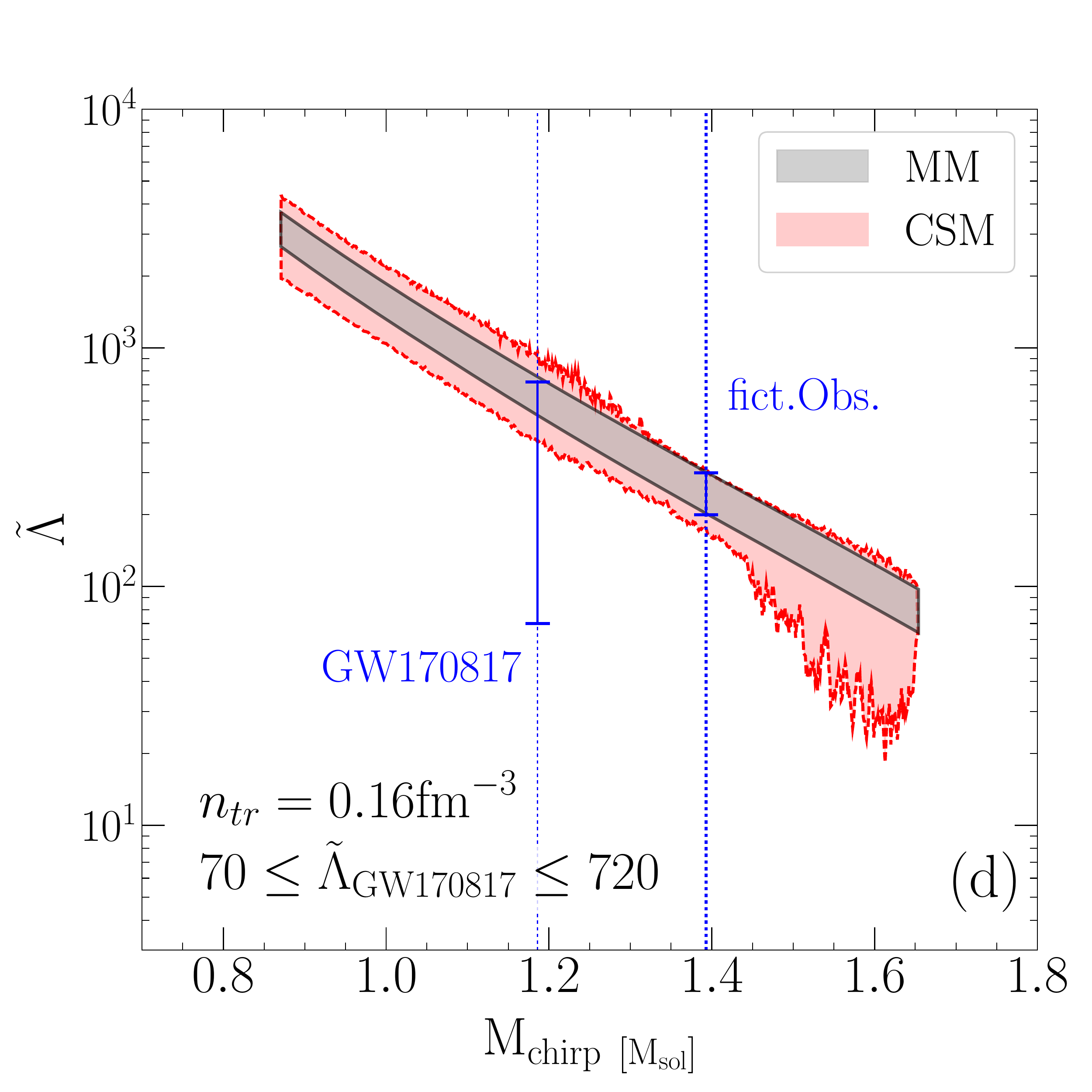}
\includegraphics[trim= 0.0cm 0 0 0, clip=,width=0.67\columnwidth]{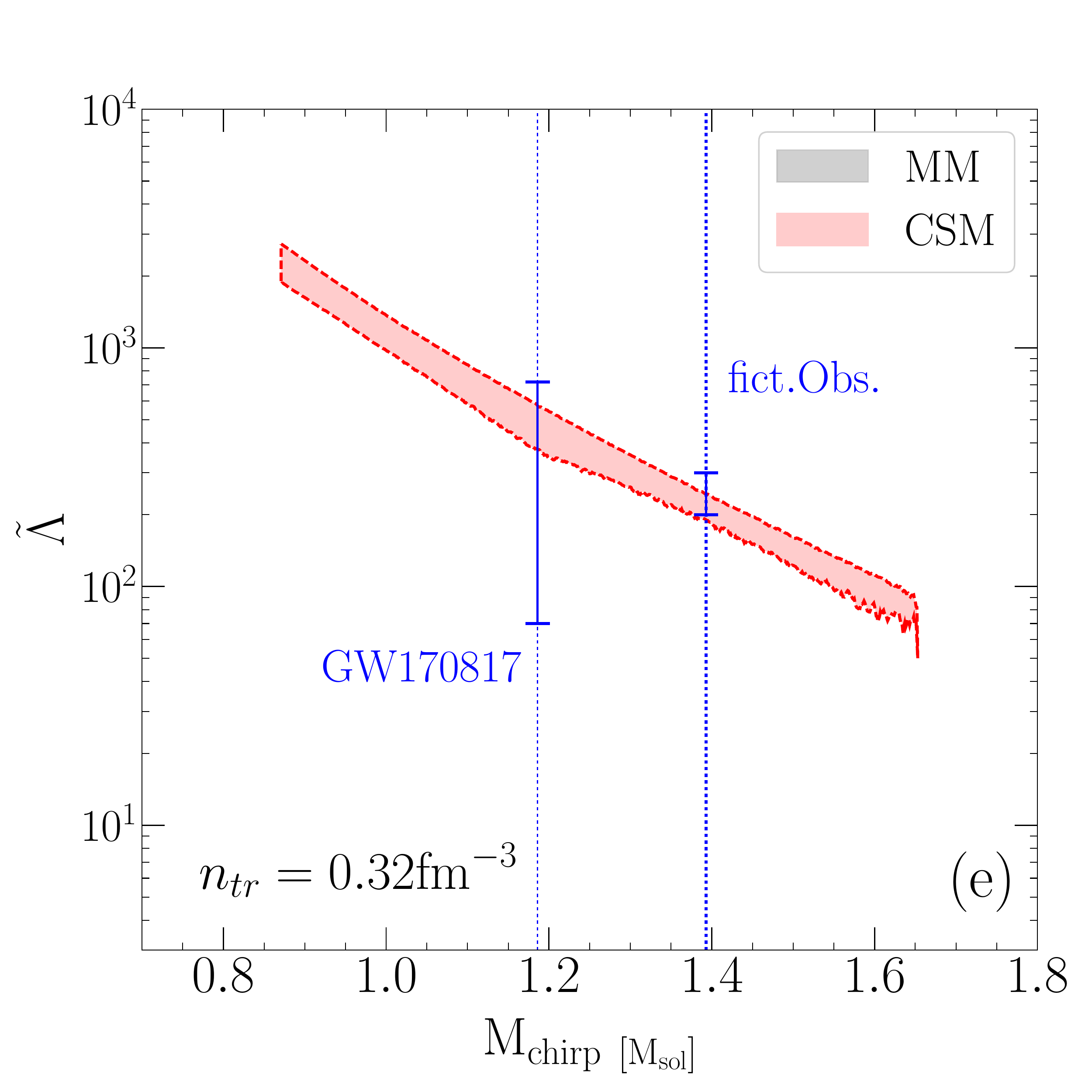}
\caption{\label{fig:MchirpLam}
Envelopes for the CSM (red) and the MM (black) for the predicted tidal polarizability parameter $\tilde{\Lambda}$ as a function of chirp mass for neutron-star binaries with component masses in the range $1.0-1.9 M_{\odot}$. We show: panel (a) the results for $n_{\text{tr}}=n_{\text{sat}}$, panel (b) for $n_{\text{tr}}=n_{\text{sat}}$ when additionally enforcing the LV constraint from GW170817, and panel (c) for $n_{\text{tr}}=2 n_{\text{sat}}$. 
In panels (d) and (e), we show how this band reduces under a fictitious observation of a merger of two $1.6 M_{\odot}$ neutron stars when $\tilde{\Lambda}$ would be measured to be $200-300$. We indicate GW170817 and the fictitious measurement (blue error bars) and the corresponding chirp masses (dotted vertical lines). In panel (e), the GW observations together with nuclear physics constraints would rule out the MM.}
\end{figure*}

It is interesting to look at areas of constant $\Lambda$ within the MR plane. In this case, the relation of neutron-star mass and radius is given by 
\begin{align}
M&=\left(\frac32 \frac{\Lambda}{k_2}\right)^{-\frac15} \frac{c^2}{G} R\,,
\end{align}
leading to the following scaling relation,
\begin{align}
\left(\frac{M}{M_{\odot}}\right)&=0.6243 \left(\frac{\Lambda}{k_2}\right)^{-\frac15} \left(\frac{R}{1 \km}\right)\,.
\label{eq:scaling}
\end{align}
For constant $\Lambda$, this implies an almost linear relationship between M and R, because the love number $k_2$ does not vary strongly in that case. In addition, for different values of $\Lambda$, the slopes are rather similar due to the small exponent $-1/5$. In Fig.~\ref{fig:MRLam}, we plot the mass-radius relation for $n_{\text tr}=n_{\rm sat}$ for the CSM, together with areas of constant $\Lambda$. In particular, we show areas for $\Lambda=200, 400, 800$, and $1600$.

While there is a tight correlation between radii and tidal polarizabilities, from Fig.~\ref{fig:MRLam} one can see that both quantities still provide complementary information. For example, an exact observation of the tidal polarizability of a neutron star, i.e., with vanishing uncertainty, would still lead to a remaining uncertainty for the radius of a typical $1.4 M_{\odot}$ neutron star. To be specific, for $\Lambda=200$, the remaining radius uncertainty is still $\approx 1$ km, compatible with the expected uncertainty of NICER~\cite{NICER1}. For larger values of $\Lambda$ this uncertainty decreases and for $\Lambda=800$ it is only $\approx 0.5$ km. However, based on GW170817 values larger than $720$ are ruled out for typical neutron stars. Hence, both tidal deformabilities and radii offer complementary information on neutron-star global structure. 

Finally, from Eq.~(\ref{eq:scaling}), one can infer the following fit,
\begin{align}
\left(\frac{M}{M_{\odot}}\right)&= \frac{a}{(b+\Lambda)^{1/5}} \left(\frac{R}{1 \km}\right)\,,
\label{eq:scalingFit}
\end{align}
where we find $a=0.406435$ and $b= 68.5$.

\subsection{Tidal polarizabilities of GW170817}

For neutron-star mergers, the GW signal allows the extraction of the binary tidal polarizability parameter $\tilde{\Lambda}$. This parameter is defined as a mass-weighted average of the individual tidal polarizabilities, 
\begin{equation}
\tilde{\Lambda}~=~\frac{16}{13} \left[\frac{(m_1+12m_2)m_1^4\Lambda_1 }{m_{\text{tot}}^5}+ \frac{(m_2+12m_1)m_2^4\Lambda_2 }{m_{\text{tot}}^5}\right]\,.
\end{equation}
As discussed in Sec.~\ref{sec:posterior}, the extraction of the binary tidal polarizability suffers from increased uncertainties, due to its importance only during the last few orbits~\cite{Flanagan2008,Damour2009} and correlations among the parameters. In the initial publication of the LV collaboration~\cite{Abbott:2017}, the constraint on $\tilde{\Lambda}\leq 800$ was reported with 90\% confidence (corrected to be $\tilde{\Lambda}\leq 900$ in Ref.~\cite{Abbott:2018wiz}). This analysis, however, was very general and did not assume both objects in the binary system to have the same EOS. Several reanalyses have since improved this constraint. Assuming that both compact objects were neutron stars governed by the same EOS, Ref.~\cite{De:2018uhw} used polytropic EOS models and a Bayesian parameter estimation with additional information on the source location from EM observations to derive limits on $\tilde{\Lambda}$ for different prior choices for the component masses: for uniform priors the reported 90\% confidence interval was $\tilde{\Lambda}=84-642$, for a component mass prior
informed by radio observations of Galactic double neutron stars the result was $\tilde{\Lambda}=94-698$, and for a component mass
prior informed by radio pulsars the reported result was $\tilde{\Lambda}=89-681$. A reanalysis by the LV collaboration found a new 90\% confidence of $70 \leq\tilde{\Lambda}\leq 720$~\cite{Abbott:2018wiz}; see Fig.~\ref{fig:posteriors}. Finally, the LV collaboration reported an additional result, assuming that both merging objects were neutron stars governed by the same EOS~\cite{Abbott:2018exr}. This EOS was based on the Lindblom parametrization~\cite{Lindblom:2010bb}
stitched to the SLy EOS for the crust, and resulted in $\tilde{\Lambda}=70-580$ with 90\% confidence. For the different extractions, the lower limit is rather stable, but the upper limit varies from 580-800. 

In general, the uncertainty range for all extractions is sizable. In the following, we will investigate the resulting $\tilde{\Lambda}$ obtained from state-of-the-art nuclear-physics models at low densities. To obtain these results, for all our EOS models we compute the combined tidal polarizability $\tilde{\Lambda}$ for thousands of NS-NS binaries where the sample the mass $m_1$ of the heavier neutron star in the range $1.0-1.9 M_{\odot}$ and the mass of the lighter neutron star $m_2$ in the range $1.0 M_{\odot}-m_1$ (implying $q\leq 1$). This allows us to explore a wide range of mass asymmetries and chirp masses ranging from $0.871M_{\odot}$ to $1.654 M_{\odot}$, which naturally includes the chirp masses for several known neutron-star binaries as well as GW170817.  We show the resulting envelopes for $\tilde{\Lambda}$ as a function of $M_{\rm{chirp}}$ in Fig.~\ref{fig:MchirpLam}. We also indicate the chirp mass for GW170817, $M_{\rm chirp}^{\rm GW170817}=1.186 M_{\odot}$~\cite{Abbott:2018wiz}  (blue dashed vertical lines) that allows to extract nuclear-physics constraints on $\tilde{\Lambda}$ of GW170817. 

Using nuclear-physics constraints from chiral EFT up to $n_\text{sat}$ [panel (a)] leads to the widest allowed range for $\tilde{\Lambda}$ for a given chirp mass. This is true for both the MM and the CSM, but the CSM envelope is much larger due to the wider flexibility of the EOS at higher densities. For GW170817 ($M_{\rm chirp}^{\rm GW170817}=1.186 M_{\odot}$), we find $\tilde{\Lambda}_{\text{CSM}}=60-2180$ and $\tilde{\Lambda}_{\text{MM}}=230-950$; for the CSM, the uncertainty in $\tilde{\Lambda}$ is much larger than the LV constraint for GW170817. For this transition density, both the MM and the CSM can be constrained by the LV constraint on GW170817 and, as a result, GW170817 adds information on the mass-radius relation of neutron stars. 

To explore the impact of the LV constraint of Ref.~\cite{Abbott:2018wiz}, we make use of $p(q,M_{\text{chirp}})$ and, using a uniform prior, select only EOS-$m_{1,2}$ combinations with $70\leq\tilde{\Lambda}\leq 720$. In panel (b) of Fig.~\ref{fig:MchirpLam} we show the resulting envelope for $\tilde{\Lambda}(M_{\rm{chirp}})$ for the MM and CSM. In addition, we also show the resulting envelopes for the EOS and the MR relation in panels (b) of Figs.~\ref{fig:EpsPcomp} and~\ref{fig:MRcomp}, respectively. Please note that the resulting range of tidal polarizabilities for $M_{\rm{chirp}}=1.186$ of $\tilde{\Lambda}=70-1020$ in Fig.~\ref{fig:MchirpLam}(b) is larger than the LV constraint. The reason is that we accept all EOS that fulfill the LV constraint for any value of $q$ allowed according to $p(q)$. The range in Fig.~\ref{fig:MchirpLam}(b), however, is computed for many more values of $q$. For example, if an EOS passes the constraint $\tilde{\Lambda}\leq 720$ for $q=0.7$ than the resulting $\tilde{\Lambda}$ for $q=1$ will be larger. 

Naturally, enforcing this constraint rules out a considerable part of EOSs that lie both on the high-pressure and low-pressure side at high energy densities. This, again, is reflected in the mass-radius relation, where neutron stars with large radii are excluded by the LV constraint. For our analysis and the CSM, we find that the radius of a $1.4 M_{\odot}$ neutron star, $R_{1.4}$, can be constrained to be $9.0\, \text{km}< R_{1.4}<13.6$ km. This was also found in Ref.~\cite{Annala:2017llu}, where a polytropic EOS expansion was used to constrain the radius of neutron stars by enforcing the constraint $\Lambda_{1.4}<800$ (the initial LV constraint of Ref.~\cite{Abbott:2017}). Ref.~\cite{Annala:2017llu} found that $R_{1.4}<13.6$ km, and both analyses are in excellent agreement. 

Finally, we assume the chiral EFT constraint to be valid up to $2n_\text{sat}$ [panel (c)]. Even though the uncertainties are still sizable, the predicted total range for $\tilde{\Lambda}$ reduces dramatically. For GW170817, we find $\tilde{\Lambda}_{\text{CSM}}=80-580$ and $\tilde{\Lambda}_{\text{MM}}=280-480$. Our constraint, which is solely guided by nuclear-EFT input, is much tighter than the observational LV constraint and in excellent agreement with the recent detailed reanalysis by the LV collaboration~\cite{Abbott:2018exr}. We emphasize, though, that our analysis is more constraining than the LV reanalysis: our 100\% envelopes are compatible with the 90\% contour of Ref.~\cite{Abbott:2018exr}. Therefore, the sentiment that the neutron-star merger GW170817 revolutionized our understanding of the EOS, is a bit of an exaggeration. GW170817, however, represents a new hope for obtaining different constraints on the EOS that might also offer the possibility to investigate new phases of dense matter. In this sense, GW170817 and the expected future detections will surely contribute to answering the long standing question of the nature of the NS core.

We explicitly stress that our results imply that current nuclear physics knowledge in the relevant density range of $1-2 n_{\rm sat}$, as obtained by ab inito calculations using modern nuclear Hamiltonians and state-of-the art many-body methods, is compatible with the recent neutron-star merger observation but more constraining for neutron-star observables and the EOS. In addition, efforts in the nuclear-theory community to improve nuclear interactions might allow to considerably reduce the theoretical uncertainty for the neutron-star-matter EOS between $1-2 n_{\rm sat}$, which will tighten our constraints even more. In general, this very interesting density range provides an excellent laboratory to probe nuclear-theory predictions against astrophysical observations and heavy-ion collision experiments.

\subsection{Impact of varying ${\bf n_{\text{tr}}}$ and the validity of chiral EFT predictions}

\begin{figure*}[t]
\begin{center}
\includegraphics[trim= 0.0cm 0 0 0, clip=,width=0.9\columnwidth]{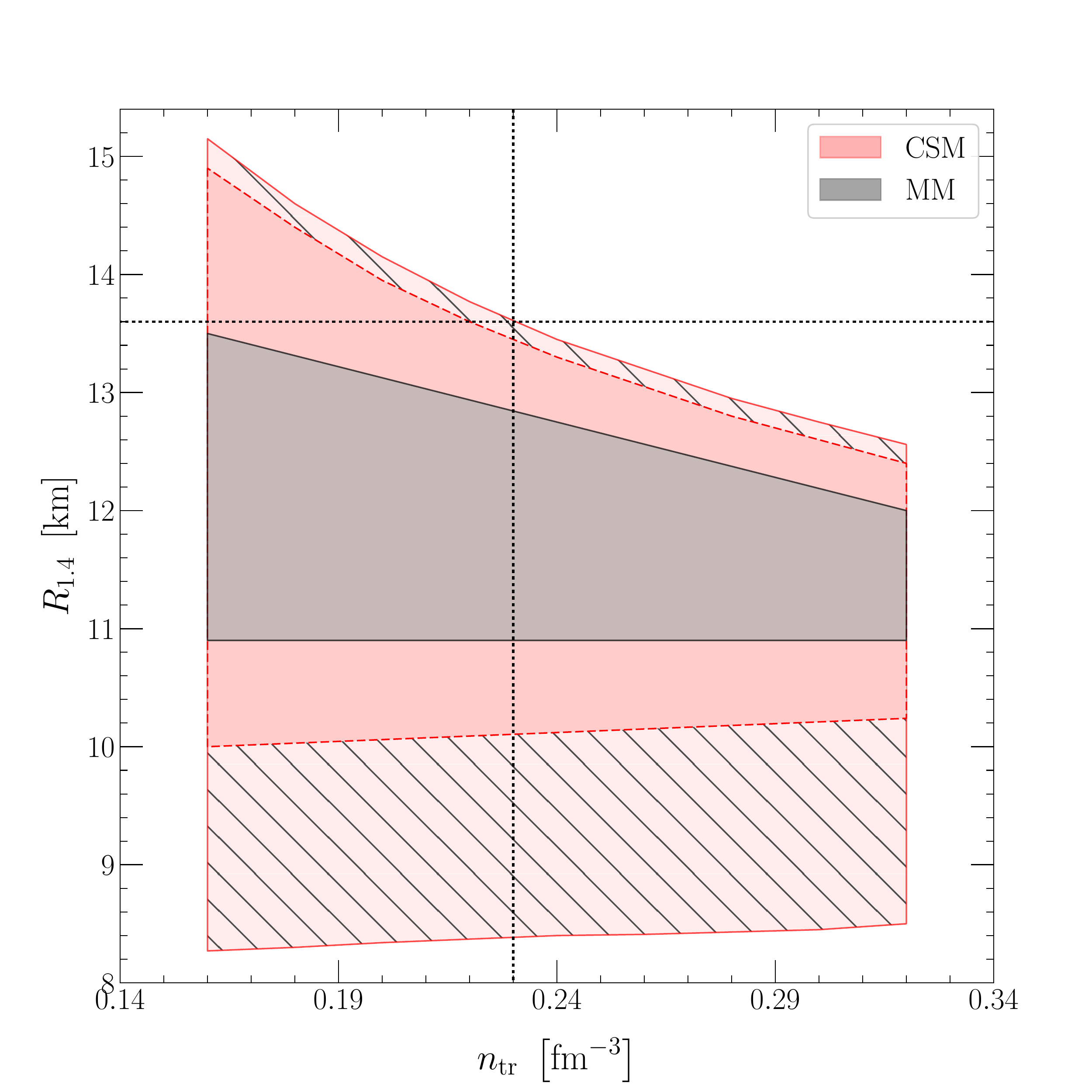}
\includegraphics[trim= 0.0cm 0 0 0, clip=,width=0.9\columnwidth]{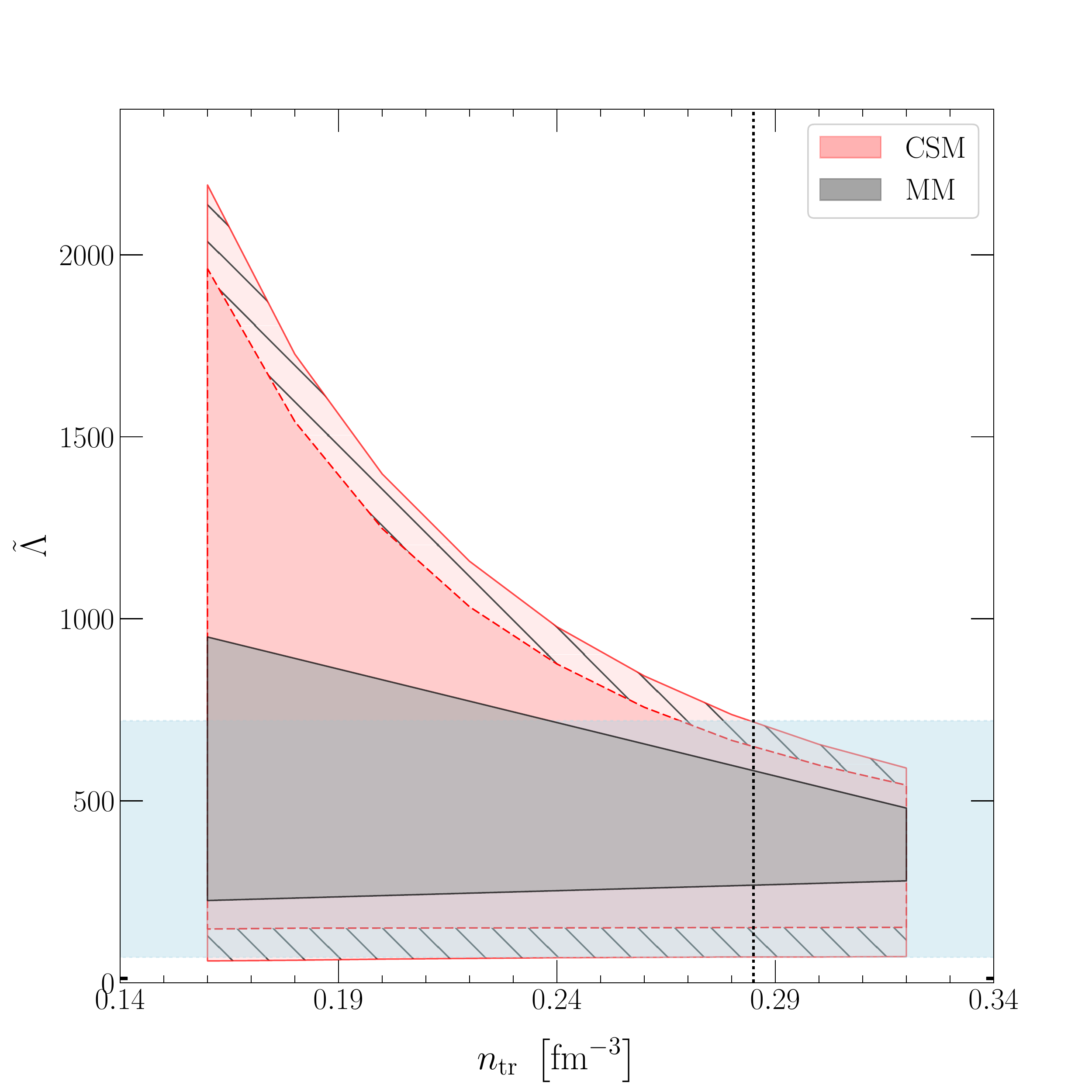}
\end{center}
\caption{\label{fig:ntrRminRmax}
Radius of a typical $1.4 M_{\odot}$ neutron star, $R_{1.4}$ (left), and $\tilde{\Lambda}$ for $M_{\rm chirp}=1.186 M_{\odot}$ (right) as functions of $n_{\rm{tr}}$. We show the envelopes for the CSM in red and for the MM in black. For the CSM, when requiring $c_S^2\leq 0.5$ instead of $c_S^2\leq1.0$, the hatched areas are excluded. We also indicate the constraints from GW170817 and the values of $n_{\rm{tr}}$, above which nuclear-theory input alone becomes more constraining than observations.}
\end{figure*} 

These present studies as well as the one of Ref.~\cite{Tews:2018kmu} are the first to use chiral EFT calculations of the neutron matter EOS up to twice nuclear saturation density with reliable error estimates to compute tidal polarizabilities for GW170817. Reliable uncertainty estimates are critical for understanding the impact that GW detections will have on elucidating the properties of dense matter inside neutron stars, and theoretical calculations of the dense-matter EOS without uncertainty estimates are of limited value for a meaningful analysis of GW data. Uncertainty estimates have shown that chiral EFT input remains useful up to $2 n_{\rm sat}$, and we find, in contrast to other recent publications~\cite{Annala:2017llu,Fattoyev:2017jql,Most:2018hfd}, that GW170817 does \emph{not} provide new insight about the EOS that cannot be obtained from current nuclear physics knowledge. This message tempers claims made in these recent publications which state that the upper limit on the tidal polarizability derived from GW data rules out stiff nuclear EOS. While this inference is correct, such stiff EOSs are already ruled out based on state-of-the-art nuclear Hamiltonians. In other words, models of dense matter excluded by the upper limit on the tidal deformability from GW170817 are already incompatible with the current microscopic EOSs at densities where error estimates can still be justified. 

Nevertheless, the reliability of chiral interactions at these densities has been questioned. Although the convergence of the chiral expansion cannot be strictly proven in this density range, we present arguments to show that the order-by-order convergence of the chiral expansion for the EOS up to $2n_{\rm sat}$ is still reasonable. First, the expansion parameter increases by only about 25\% over the density interval $1-2 n_{\rm sat}$. Second, Ref.~\cite{Tews:2018kmu} analyzed the order-by-order convergence of the employed Hamiltonians at $2 n_{\rm sat}$, and showed that, even though the reliability naturally decreases with increasing density, the order-by-order convergence remains reasonable and consistent with simple power counting arguments within the theoretical uncertainty estimates. Nevertheless, densities around $2 n_{\rm sat}$ seem to provide an upper limit to the applicability of the chiral Hamiltonians we use in this work.

To support our main statement - namely that the constraints from GW170817 are compatible with but less restrictive than predictions of the EOS based on realistic nuclear potentials and do not yield specific new information about nuclear Hamiltonians or about possible phase transitions at supra-nuclear density - in this context, we investigate which density range for chiral EFT input is sufficient to justify our statement. We present the total uncertainty ranges for $R_{1.4}$ (left panel) and $\tilde{\Lambda}$ for $M_{\rm chirp}=1.186 M_{\odot}$(right panel) as functions of the density $n_{\rm tr}$ in Fig.~\ref{fig:ntrRminRmax}. For $R_{1.4}$, we indicate the upper limit on the radii of Ref.~\cite{Annala:2017llu}, $R_{1.4}\leq 13.6$ km, which was obtained using $n_{\rm tr}=n_{\rm sat}$ and the LV constraint (horizontal dotted line). We find that the CSM alone constrains the radii to be smaller than this bound for $n_{\rm tr}>0.23 \fmiq \approx 1.44 n_{\rm sat}$ (an 11\% increase of the expansion parameter compared to $n_{\rm sat}$). For the tidal polarizability, we indicate the LV constraint as a horizontal blue band and find that the CSM leads to $\tilde{\Lambda}\leq 720$ as soon as $n_{\rm tr}> 0.285 \fmiq \approx 1.78 n_{\rm sat}$ (a 20\% increase of the expansion parameter compared to $n_{\rm sat}$). We would like to emphasize that these crucial values for $n_{\rm tr}$ for both observables do not necessarily have to agree, as seen in Fig.~\ref{fig:ntrRminRmax}. The reason is that the upper limit on $\tilde{\Lambda}$ depends on $q$ while $R_{1.4}$ does not. In Fig.~\ref{fig:MchirpLam}(b) we have seen that when varying $q$ in the range allowed by GW170817, $\tilde{\Lambda}$ can increase to values $\sim 1000$ for the EOS that pass the LV constraint from GW170817. Chiral EFT input becomes compatible with this value at $n_{\rm tr}\sim 0.23 \fmiq$, in agreement with the value for $R_{1.4}$. At these values for $n_{\rm tr}$, in particular at $1.44 n_{\rm sat}$, arguments for the validity of chiral interactions remain even stronger,  which strengthens the validity of our main statement.

Finally, the value of $n_{\rm tr}$ also affects the speed of sound inside neutron stars. The speed of sound is expected to approach the conformal limit of $c_S^2=1/3$ at very high densities~\cite{Kurkela:2010}. In neutron stars, though, it is not clear if this conformal limit remains valid or not. As discussed in detail in Ref.~\cite{Tews:2018kmu}, the neutron-matter EOS up to $n_{\rm tr}=2 n_{\rm sat}$ requires the speed of sound to pass the conformal limit to be sufficiently stiff to stabilize the observed two-solar-mass neutron stars. In fact, for chiral models the speed of sound has to increase beyond the conformal limit for $n_{\rm tr}>0.28 \fmiq$ and even for phenomenological nuclear Hamiltonians, which lead to stiffer neutron-matter EOS, this statement remains valid for $n_{\rm tr}>0.31 \fmiq$. While there might be EOS that are much stiffer below $2 n_{\rm sat}$ and, hence, stabilize the heaviest neutron stars while still obeying the conformal limit, such EOS are ruled out by modern nuclear Hamiltonians. 

Therefore, the neutron-matter EOS up to $2 n_{\rm sat}$ for state-of-the-art nuclear Hamiltonians requires the speed of sound in neutron stars to experience a non-monotonous behavior, i.e, increasing beyond $c_S^2=1/3$ but decreasing at higher densities to approach this limit.
For example, for chiral EFT interactions and $n_{\rm tr}=2 n_{\rm sat}$, the speed of sound has to reach values $c_S^2\geq 0.4$. 
The question remains, though, which forms of strongly-interacting matter lead to such a behavior for the speed of sound. 
In order to estimate the impact of the speed-of-sound behavior on $R_{1.4}$ and $\tilde{\Lambda}$, we present hatched areas in Fig.~\ref{fig:ntrRminRmax} which are excluded for $c_s^2\leq0.5$. We choose this limiting value solely for illustrative purposes.
This constraint slightly reduces the upper bound on neutron-star radii but it would mostly rule out low-radius neutron stars.  
The reason is that neutron stars can have very small radii only for strong first-order phase transitions with low onset densities. To simultaneously support $2M_{\odot}$ neutron stars, the EOSs has to experience a sudden subsequent stiffening, i.e., the speed of sound has to increase dramatically. For a larger possible speed of sound, stronger phase transitions are allowed, which leads to stars with smaller radii. Limits on $c_S^2$, on the other hand, rule out the strongest phase transitions, and increase the smallest possible radius. For $c_S^2\leq 0.5$, the lower limit on the radius of a $1.4M_{\odot}$ neutron star is approximately 10 km, of the order of the constraint of Ref.~\cite{Bauswein:2017vtn}.

\subsection{Impact of additional constraints}

\begin{figure}[t]
\includegraphics[trim= 0.0cm 0 0 0, clip=,width=0.9\columnwidth]{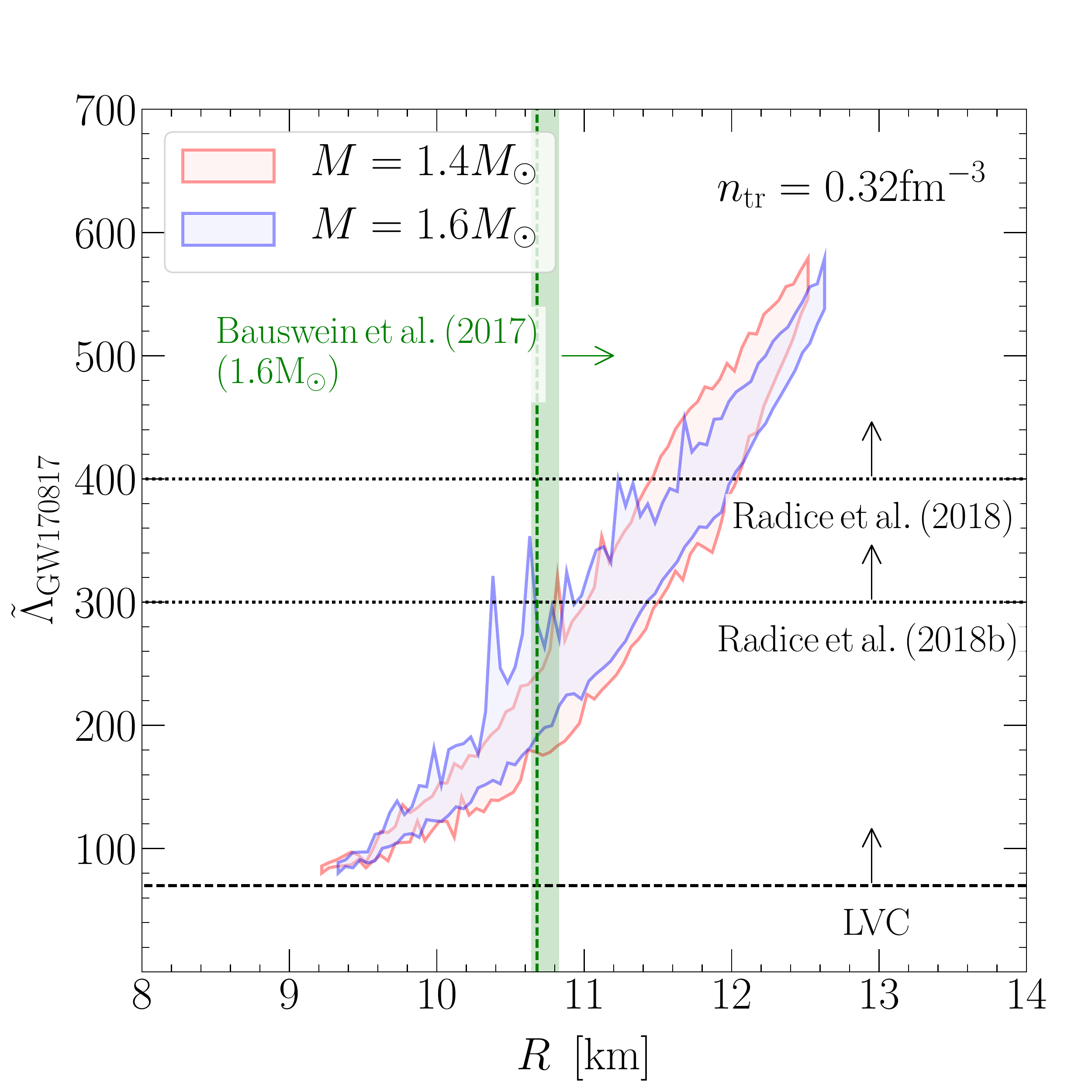}
\caption{\label{fig:RLam}
Envelopes for the correlation between $\tilde{\Lambda}$ of GW170817 and the radius of a $1.4 M_{\odot}$ (red) and the radius of a $1.6 M_{\odot}$ (blue) neutron star for $n_{\text{tr}}=2 n_{\text{sat}}$ and the CSM. The corresponding values for the MM (not shown) lie within the CSM envelopes. 
We also show the lower limit of the LV constraint on the tidal polarizability of GW170817~\cite{Abbott:2018wiz}, the proposed constraint of Ref.~\cite{Radice:2017lry} and its update of Ref.~\cite{Radice:2018ozg}, and the radius constraint for a $1.6 M_{\odot}$ neutron star from Ref.~\cite{Bauswein:2017vtn}. 
}
\end{figure} 

Even though the tidal polarizabilities extracted from GW170817 alone may not revolutionize our understanding of the EOS, several additional constraints based on the EM counterpart were proposed. These additional constraints were mostly based on the fact that the EM signal of GW170817 does not seem to imply a prompt collapse of the hypermassive merger remnant to a black hole. Instead, it is argued that the merger remnant survived for several 100 milliseconds before collapse. Based on this assumption, several groups independently suggested the maximum mass of neutron stars to be less than $\approx 2.2-2.3 M_{\odot}$~\cite{Margalit:2017,Shibata:2017xdx,Rezzolla:2017aly}. While this constraint is powerful for smooth EOS models, which exhibit a strong correlation between $M_{\rm max}$ and radii of typical neutron stars, the appearance of strong first-order phase transitions in general EOS models implies that the maximum mass is not very constraining for the structure of typical neutron stars; see also Ref.~\cite{Tews:2018iwm}. 

Additional constraints for radii and tidal polarizabilities were proposed based on the same assumptions. Ref.~\cite{Bauswein:2017vtn} suggested that the EM observation can be used to argue that $R_{1.6}\geq 10.68_{-0.04}^{+0.15}$ km. In contrast to the $M_{\rm max}$ constraint, a radius constraint has a sizable impact on the CSM: In Figs.~\ref{fig:EpsPcomp}(b) and (c) as well as Figs.~\ref{fig:MRcomp}(b) and (c) we indicate parts of the envelopes which are excluded by $R_{1.6}\geq 10.68_{-0.04}^{+0.15}$ km by hatched areas. In addition,  Ref.~\cite{Radice:2017lry} suggested that the amount of ejecta determined from the EM observations implies $\tilde{\Lambda}>400$. This constraint was later updated to $\tilde{\Lambda}>300$~\cite{Radice:2018ozg}.  In Fig.~\ref{fig:RLam}, we show the correlation between $\tilde{\Lambda}$ and the radii of a $1.4 M_{\odot}$ neutron star, $R_{1.4}$, and a $1.6 M_{\odot}$ neutron star, $R_{1.6}$, for $n_\text{tr}=2n_\text{sat}$ and the CSM. While in general radius and tidal polarizabilities are correlated, the appearance of phase transitions washes this correlation out. Fig.~\ref{fig:RLam} again highlights the fact that even an exact determination of $\tilde{\Lambda}$ leaves a considerable radius uncertainty. Therefore, independent observations of radii and tidal polarizabilities are crucial to pin down the high-density EOS of nuclear matter.

\begin{figure*}[t]
\begin{center}
\includegraphics[trim= 0.0cm 0 0 0, clip=,width=0.9\columnwidth]{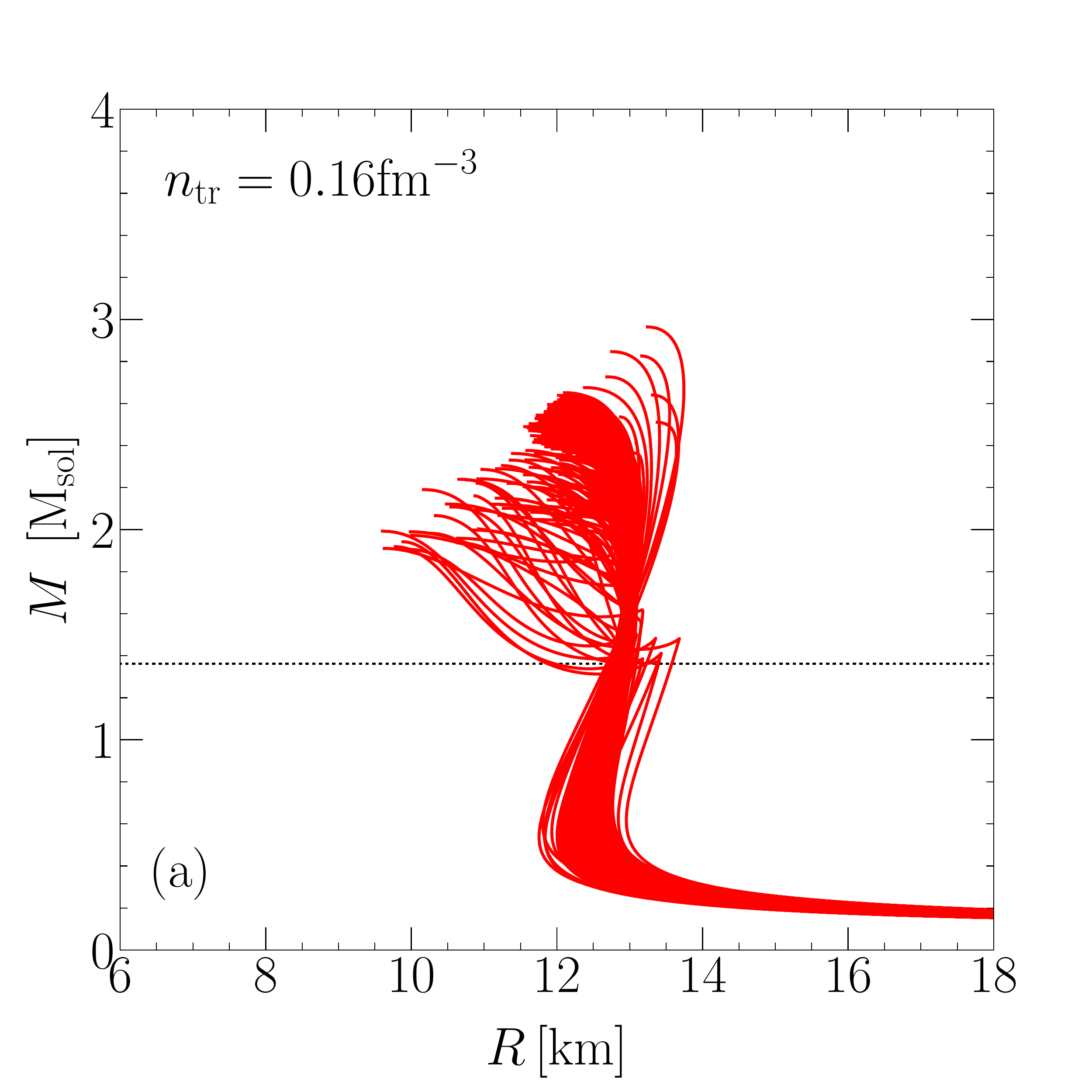}
\includegraphics[trim= 0.0cm 0 0 0, clip=,width=0.9\columnwidth]{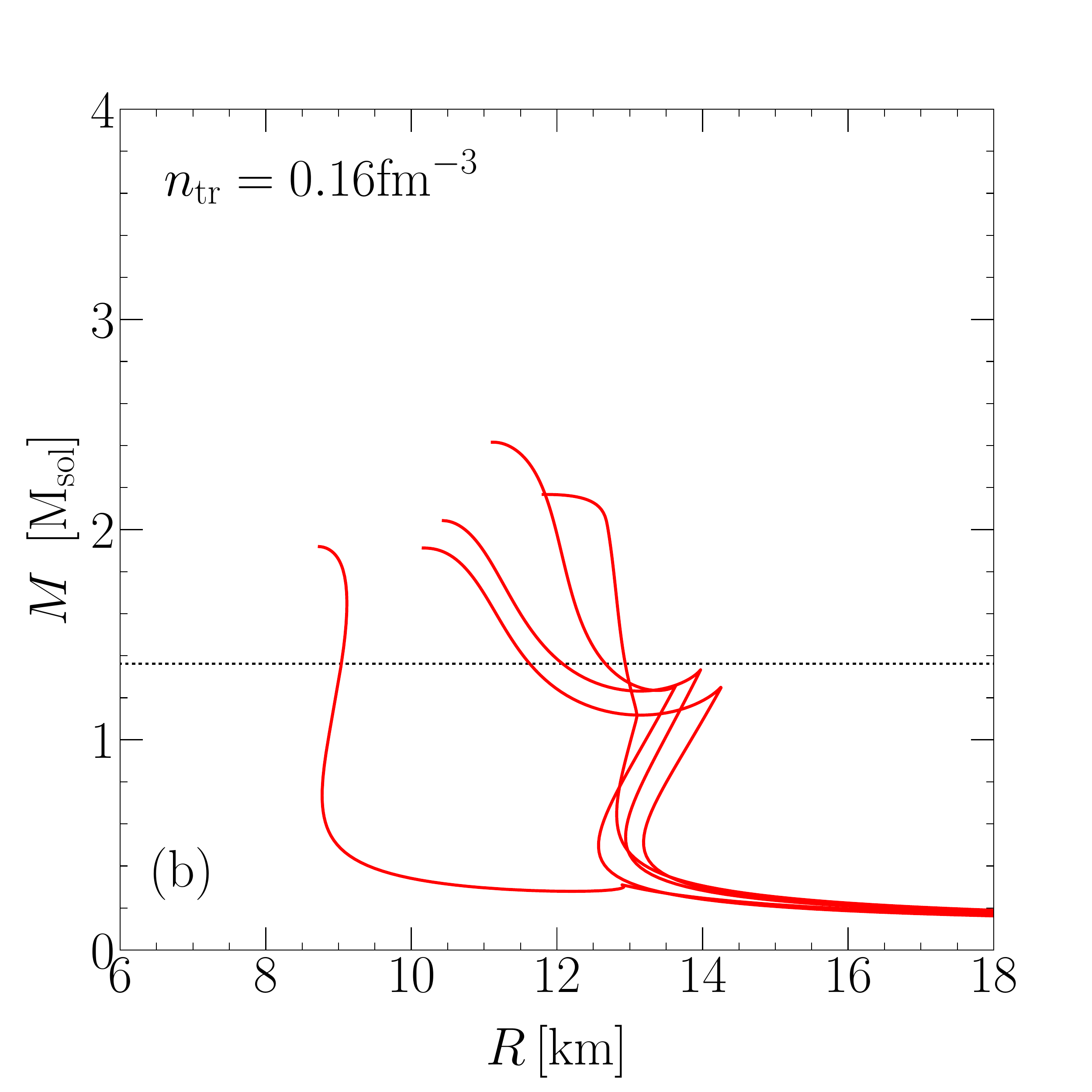}
\end{center}
\caption{\label{fig:q0710diff}
Equations of state for $n_{\rm tr}=n_{\rm sat}$ which pass the LV constraint $70\leq \tilde{\Lambda}\leq 720$ for $q=0.7$ but not for $q=1.0$ [panel (a)] and vice versa [panel (b)].
}
\end{figure*} 

In Fig.~\ref{fig:RLam}, we also show the constraints of Refs.~\cite{Bauswein:2017vtn,Radice:2017lry,Radice:2018ozg}. The radius constraint implies that $\tilde{\Lambda}\geq 180$ while the constraint of Ref.~\cite{Radice:2017lry} (Ref.~\cite{Radice:2018ozg}) implies $R_{1.6} \sim R_{1.4}\geq 11.5$ km ($10.5$ km). All of these constraints are based on empirical formulas extracted from simulations for a limited set of model EOSs. Especially for the constraints of Refs.~\cite{Radice:2017lry,Radice:2018ozg}, this set contains only four nucleonic EOS and, therefore, is likely overestimated~\cite{Tews:2018iwm}. 
In the case of the first constraint, a similar argument might be made. Nevertheless, in that case the authors try to explore the full EOS dependence which results in a more conservative constraint.
In both cases, however, future numerical simulations with additional EOSs, including, e.g., phase transitions, can be used to refine these constraints and improve their robustness.

In addition to inferences from GW170817, additional future observations might dramatically improve our understanding of the EOS.  
The NICER~\cite{NICER1} and eXTP~\cite{Watts:2018iom} missions will provide neutron-star radii with a few percent uncertainty: the NICER mission is expected to provide first results within this year. As we have seen above, these future radius observations might considerably reduce the ambiguity of the allowed EOS models. A measurement of $R_{1.4}$ with a 5\% accuracy will add valuable information and might help distinguish EOSs with and without phase transitions; see also Ref.~\cite{Tews:2018kmu}.

In addition, in the next years additional neutron-star merger observations by the LV collaboration are expected. While the uncertainty for the tidal polarizability associated with GW170817 is not sufficient to constrain the EOS, this might change for future observations. For example, mergers with better signal-to-noise ratios could be observed, or sufficiently many mergers are observed so that accurate information can be extracted. In addition, third generation GW detectors might provide tidal-polarizability measurements with 10\% uncertainty. To illustrate the possibilities offered by such new GW events, we inject in Fig.~\ref{fig:MchirpLam}(d) and (e) a fictituous measurement of $M_{chirp}=1.385$ and $\tilde{\Lambda}$ to be measured in the range $200-300$. Such an observation would dramatically reduce the uncertainties in the EOS: 
it would reduce the allowed radius range for a typical neutron star to 
11.7-13.4 km for $n_{\rm tr}=n_{\rm sat}$ and to only 11.7-12.5 km for $n_{\rm tr}=2 n_{\rm sat}$. Also, it is interesting to note that in this case the MM cannot reproduce the two events, GW170817 and the fictitious one. There is, therefore, a great potential to combine future detections as a filter for EOS models and the accumulation of GW tidal deformabilities may offer the possibility to make statements about the existence of phase transitions in dense matter.

 \begin{figure*}[t]
\begin{center}
\includegraphics[trim= 0.0cm 0 0 0, clip=,width=0.9\columnwidth]{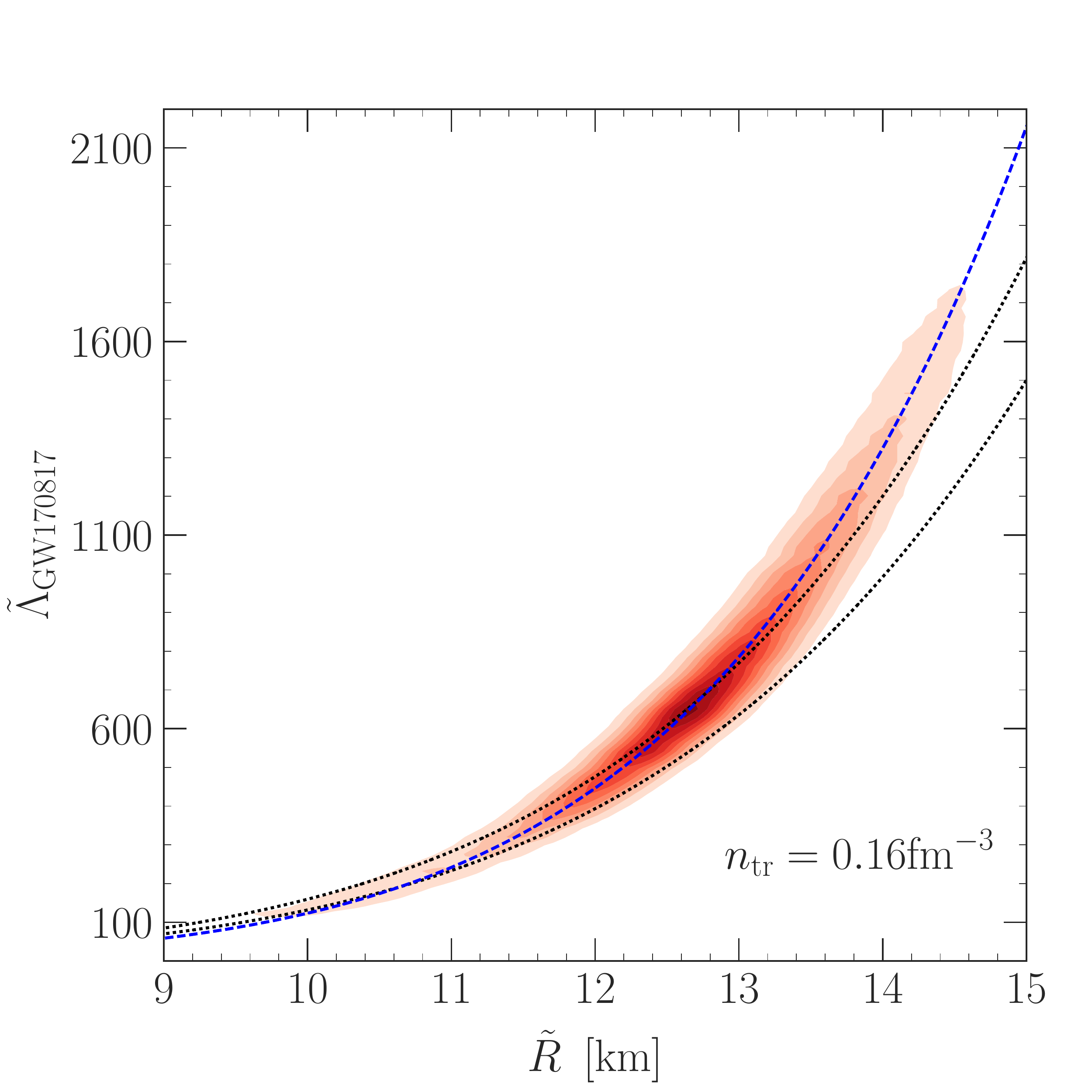}
\includegraphics[trim= 0.0cm 0 0 0, clip=,width=0.9\columnwidth]{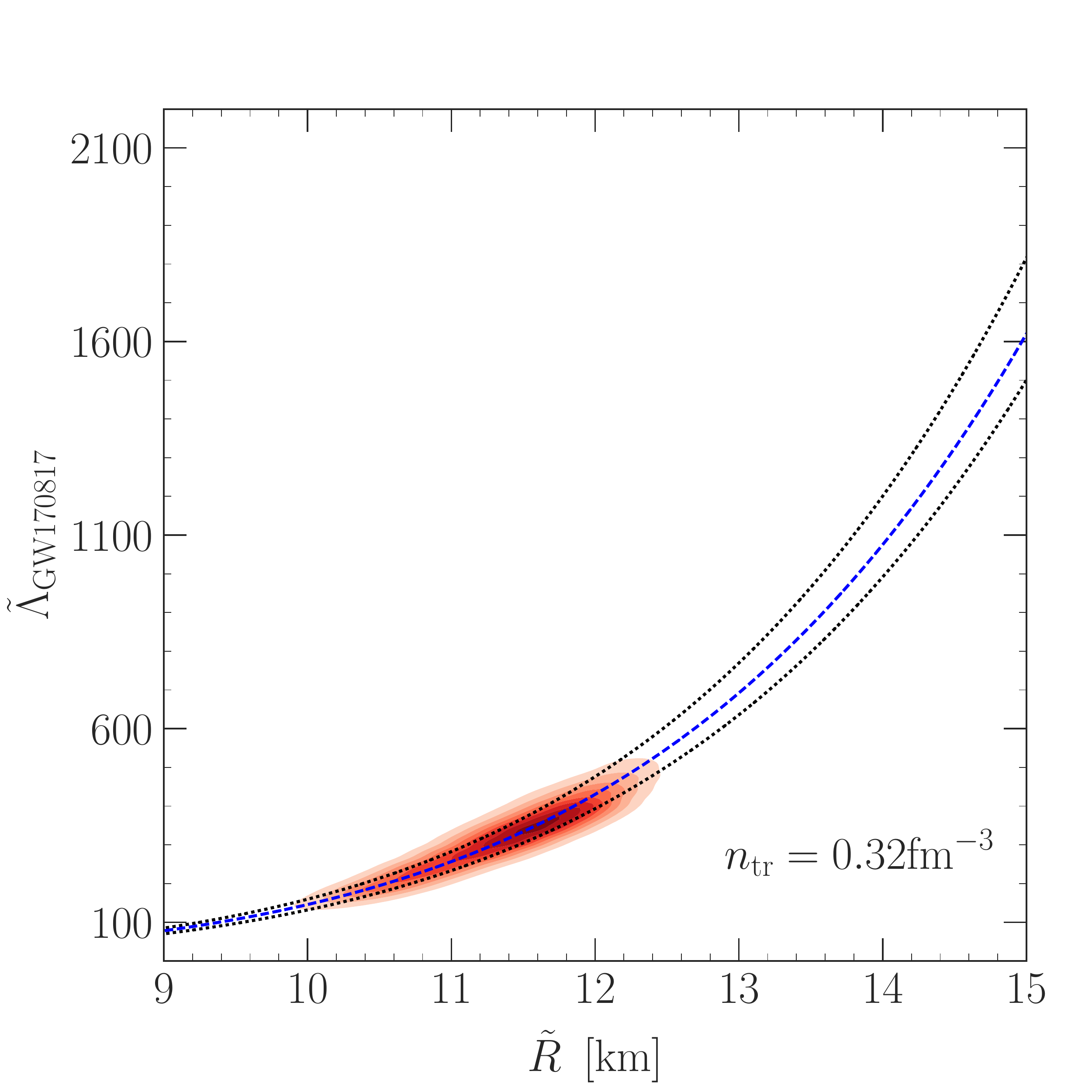}
\end{center}
\caption{\label{fig:EmpRel}
Relation connecting the common radius $\hat{R}$  and the binary tidal polarizability $\tilde{\Lambda}$ for $0.7<q<1.0$ and for $n_{\rm tr}=n_{\rm sat}$ (left panel) or $n_{\rm tr}=2 n_{\rm sat}$ (right panel). As a comparison, we show the relation Eq.~(5) of Ref.~\cite{De:2018uhw}
with its uncertainty (black dotted lines) and our fits (blue dashed line).}
\end{figure*} 

\subsection{Impact of phase transitions on tidal polarizability}
\label{sec:PTimpact}

In the previous sections, we have seen that ranges for all neutron-star observables are larger for the CSM than the MM because the CSM permits regions of drastic stiffening or softenting of the EOS. In this section, we briefly discuss the impact that strong phase transitions have on neutron-star tidal polarizabilities. 

Of special interest for the interpretation of merger observations is the behavior of the EOS for stars in the mass range of the two component masses: For GW170817 this range is around $M=1.4M_{\odot}$. EOS with strong first-order phase transitions appearing in stars of this mass range might be probed by future merger observations. For instance, the CSM, which includes such phase transitions, permits small values for $\tilde{\Lambda}$ due to strong softening and subsequent stiffening of the EOS, but the MM prevents $\tilde{\Lambda}$ to be below $\approx 250$. These observable differences among the two models allow us to identify ranges of tidal deformabilities (and neutron-star observables in general) for which a strong first-order phase transition is preferred or even necessary, providing a means to probe new states of matter inside neutron stars. In the above example, an observation of $\tilde{\Lambda}\leq 250$ would indicate a softening of the EOS that smooth (nucleonic) EOS cannot provide.

We have also seen before that strong phase transitions weaken the correlation between $R$ and $\tilde{\Lambda}$. For EOSs with phase transitions in the relevant mass range, which produce lighter stars with larger radii and heavier stars with smaller radii, a significant mass asymmetry of a merging binary keeps the EOS compatible with a constraint on $\tilde{\Lambda}$ but permits larger radii for typical neutron stars and, therefore, washes out this correlation.

In Fig.~\ref{fig:q0710diff}, we illustrate this behavior for $n_{\rm tr}=n_{\rm sat}$ for two interesting cases: EOSs which pass the LV constraint for $q=0.7$ but are excluded for $q=1.0$ and vice versa. We show the EOSs belonging to the first class of models in Fig.~\ref{fig:q0710diff}(a) and the EOSs belonging to the second class of models in Fig.~\ref{fig:q0710diff}(b). In general, for a given EOS, heavier neutron stars have smaller tidal polarizabilities, and increasing the mass asymmetry in the binary, i.e., lowering $q$, results in slightly smaller values for $\tilde{\Lambda}$ for a given chirp mass. Therefore, several smooth EOSs, i.e., without phase transitions, pass the LV constraint for $q=0.7$ but not for $q=1.0$, which can be seen in Fig.~\ref{fig:q0710diff}(a).

The more interesting case are EOS models with a strong phase transition occurring around $1.4 M_{\odot}$ and leading to a kink in the MR curve. Below the kink, radii and tidal polarizabilities are larger but drastically decrease beyond the phase transition. Two cases can be distinguished: the phase transition appears at masses above $1.4 M_{\odot}$ or below $1.4 M_{\odot}$. For the first case, $q=1$ for GW170817 implies that both stars have the same mass $\sim 1.4 M_{\odot}$ and, therefore, larger radii and tidal polarizabilities $\tilde{\Lambda}=\Lambda_1=\Lambda_2$. Lowering $q$, so that the heavier star probes the phase transition, suddenly decreases $\tilde{\Lambda}$ by a fair amount. Therefore, some EOS will be rejected for $q=1$ but accepted for lower $q$, e.g., $q=0.7$. We show these models in Fig.~\ref{fig:q0710diff}(a).  In contrast to the smooth models, though, these models permit much larger radii for typical neutron stars, which can also be seen in Fig.~\ref{fig:MRcomp}(b).

If the phase transition appears below $1.4 M_{\odot}$, the inverted situation can appear: EOSs are ruled out for $q=0.7$ but allowed for $q=1.0$. We show these cases in the right panel of Fig.~\ref{fig:q0710diff}. If the phase transition happens in very low-mass stars at densities close to saturation density, then the EOS produces neutron stars with very small radii of the order of $R_{1.4}\sim 9$ km. In this case, $\tilde{\Lambda}$ is reduced for smaller values of $q$ and the EOS is ruled out due to the lower constraint the tidal polarizability, $70\leq \tilde{\Lambda}$. However, this is an extremely rare situation and we find only one such EOS among tens of thousands of samples, see Fig.~\ref{fig:q0710diff}(b). 
If the phase transition appears in stars slightly below $1.4 M_{\odot}$, for $q=1$ both stars in GW170817 would have been hybrid stars and the $\tilde{\Lambda}$ would have been small enough for these models to pass the constraint. Increasing the mass asymmetry, $\Lambda_1$ decreases but $\Lambda_2$ increases rapidly, leading to the EOS being rejected by the upper constraint on $\tilde{\Lambda}$. We found a few such models, see Fig.~\ref{fig:q0710diff}(b). 

In any case, information on possible strong first-order phase transitions might be obtained by neutron-star merger observations. The observation of two mergers with similar chirp mass but different mass asymmetries and dramatically different binary tidal polarizabilities might shed light on the location of a strong first-order phase transition. In addition, future observations accessing regions allowed by the CSM but forbidden by the MM might also provide information on such a phase transition. For these extractions, however, higher-order GW parameters need be constrained much more precisely in future observations.  

\subsection{Empirical relations for $\bf \tilde{\Lambda}$} 
\label{sec:Relations}

Finally, we use our EOS models to investigate the empirical relation between the tidal polarizability and the radius of neutron stars.  Such a relation was reported in Eq. (5) of Ref.~\cite{De:2018uhw}, that related the binary tidal polarizability $\tilde{\Lambda}$ to the common radius of a neutron-star binary:
\begin{equation}
\tilde{\Lambda}=0.0042(4) \left(\frac{\hat{R}c^2}{G M_{\rm{chirp}}} \right)^6= 0.000146(13) \left(\frac{\hat{R}}{km} \right)^6\,.
\end{equation}
Similarly, a relation between the tidal polarizability and radius of a typical $1.4 M_{\odot}$ neutron star was reported in Ref.~\cite{Annala:2017llu}:
\begin{equation}
\Lambda_{1.4}=2.88\cdot 10^{-6}\left(\frac{R_{1.4}}{km} \right)^{7.5}\,.
\end{equation}
Interestingly, even though both approaches are based on a piecewise polytropic expansion for the EOS, the resulting relations and especially exponents are rather different (for $q=1$, $\tilde{\Lambda}\sim \Lambda_{1.4}$ and $\hat{R} \sim R_{1.4}$).

We constructed similar relations between $\tilde{\Lambda}$ and the average radius of the two binary neutron stars in GW170817 for the CSM
and $n_{\rm tr}=n_{\rm sat}$ and $n_{\rm tr}=2 n_{\rm sat}$. We show density plots for our data points and the resulting fit functions in Fig.~\ref{fig:EmpRel}, together with the result of Ref.~\cite{De:2018uhw}. For $n_{\rm tr}=n_{\rm sat}$ (left panel), we find the relation 
 \begin{equation}
\tilde{\Lambda}=0.00057(6) \left(\frac{\hat{R}c^2}{G M_{\rm{chirp}}} \right)^{7.05}\,.
 \end{equation}
In this case, the exponent lies in between the other two determinations but is closer to the result of Ref.~\cite{Annala:2017llu}.  For $n_{\rm tr}=2 n_{\rm sat}$, we find instead
 \begin{align}
\tilde{\Lambda}&=0.0047(8) \left(\frac{\hat{R}c^2}{G M_{\rm{chirp}}} \right)^{5.94}\,,
\end{align}
in very good agreement to the relation of Ref.~\cite{De:2018uhw}. Comparing the findings, we see that these relations are not universal but depend on the EOS input used. 

\subsection{Comparisons to other recent works}

There is general consensus that the upper bound on the tidal deformability $\Lambda_\text{1.4}<800$ derived by the initial analysis by the LIGO-Virgo scientific collaboration in Ref.\cite{Abbott:2017} implies that the radius $R_\text{1.4}\lesssim 13.6$ km. 
Making the reasonable assumption that both compact objects were NSs, and that they are both described by the same EOS, other authors have discussed how the bound on the tidal deformability impacts our understanding of NSs and dense matter.  In what follows we compare our analysis to some of these studies. 

In Ref.~\cite{Annala:2017llu} the authors construct a model for the EOS based on the predictions of  
chiral EFT up to a baryon number density $n_{sat}$ and use a set of four polytropes to describe matter at higher densities encountered in the core. They claim that perturbative calculations of QCD (pQCD) valid at very high density, far exceeding those encountered inside the NS core, can constrain the allowed parameter space of the polytropic EOSs.  This is then combined with the upper limit on the tidal deformability to constrain the relationship between mass and radius of all NSs and the EOS of matter encountered in their cores.  The maximal model we employ addresses the question of how improved constraints on the EOS from theory between  $n_{sat}$ and $2n_{sat}$ will alter the situation. We find no evidence for the usefulness of constraints from pQCD. The pressure in NS cores is much smaller than those encountered at the densities where pQCD is valid. Our maximal model is thermodynamically consistent and has adequate freedom to satisfy constraints from pQCD, but is uninformed by it. 

In Ref. \cite{Fattoyev:2017jql} the authors use a model EOS for neutron-rich matter that describes matter at sub-nuclear density encountered inside nuclei and at higher densities encountered inside neutron stars.  They find a strong correlation between  the neutron-skin thickness of neutron-rich nuclei and the neutron star tidal deformability, similar to the correlation between the skin-thickness and neutron-star radii found earlier \cite{Horowitz:2001}.  Such a correlation is expected because the NS radius and the tidal deformability are tightly correlated in models that do not contain phase transitions. For their models they report a tight correlation given by $\Lambda \simeq 7.76 \times 10^{-4}~ (R/\text{km})^{5.3}$. Using the correlation between neutron skin thickness and NS radius they show that the experimental lower bound on the neutron-skin thickness of $^{208}$Pb implies $R_\text{1.4}>12.55$ km. This, combined with the correlation between $\Lambda$ and $R$, is used to deduce that  $\Lambda_\text{1.4}>490$. As discussed earlier, both these correlations are \emph{model dependent}.  It is useful to compare these inferences to the predictions of our minimal model shown in Fig.~\ref{fig:ntrRminRmax} which assumes a smooth EOS without phase transitions, does not violate experimental data for the neutron-skin thickness of $^{208}$Pb, but can accommodate smaller values for $R_\text{1.4}$ and $\Lambda_\text{1.4}$. 

In Ref. \cite{Most:2018hfd}, the authors impose an additional constraint requiring that $M_\text{max} < 2.16~M_\odot$ and employ EOSs with and without strong first-order phase transitions to determine limits on the neutron star radius and deformability. In the absence of phase transitions they find that $12~\text{km} < R_{1.4} < 13.45~ \text{km}$ and require $\Lambda_{1.4}> 375$. This range is deduced as the $2\sigma$ interval by exploring a large suite of hadronic models. Our analysis based on the minimal model finds that smaller radii are possible. Further, we caution against using a probabilistic interpretation of the allowed ranges for  $R_{1.4}$ and  $\Lambda_{1.4}$ because it is difficult to assign likelihoods to a specific realization of the EOS. The inclusion of strong phase transitions in \cite{Most:2018hfd} allows for the existence of "twin star" solutions containing two separate stable branches of NSs. In this case, smaller values for  $R_{1.4}$ and  $\Lambda_{1.4}$ are allowed and the constraints weaken to $R_{1.4}>8.53~\text{km}$ and  $\Lambda_{1.4}>35.5$. The results obtained using the maximal model (CSM) are in good agreement with these limits. 

\section{Summary}
\label{sec:summary}

To summarize, we confronted the recent GW observation with modern nuclear-physics constraints from chiral EFT. We elaborated on our results of Ref.~\cite{Tews:2018iwm} and provided many additional results.  

In particular, we have used two different classes of models to extend QMC results with chiral EFT interactions to higher densities encountered in the core of neutron stars. We have used a minimal model, that is based on a density expansion around saturation density, and a maximal model based on a very general expansion in the speed of sound, that explores all EOSs consistent with the low-density input from chiral EFT. We used these models to study the uncertainties for the EOS and neutron-star observables for chiral EFT input up to either $n_{\rm sat}$ or $2 n_{\rm sat}$.

We used these models with input from nuclear physics up to nuclear saturation density and data from GW170817 to deduce that the radius of a typical neutron star has to be $R_{1.4}\leq 13.6$ km. If instead EFT predictions for the EOS are used up to twice nuclear saturation density we find that $\tilde{\Lambda}<580$ and $R_{1.4}\leq 12.6$ km. These smaller ranges suggest that future observations need to provide much more precise constraints to enable conclusions about the EOS or provide evidence for novel phases of matter in neutron stars.  We compared our results to other recent works, which arrived at the opposite conclusion, and discussed the robustness of our main statement. 

We studied the impact of additional constraints on our findings. Most of these additional constraints are derived from interpretations of the EM counterpart of GW170817, and provide limits on radii, tidal polarizabilities, or the maximum mass. We showed that constraints on the maximum mass do not reduce the EOS uncertainty for typical neutron stars, in contrast to radius information, which is rather valuable. We also investigated how an upper limit on the speed of sound in neutron stars affects our findings.

We finally investigated the impact of strong first-order phase transitions on our predictions. Contrasting the predictions of the MM and the CSM may provide useful insights on how future measurements of $\tilde{\Lambda}$ from neutron-star mergers can help to identify new forms of matter at densities beyond nuclear saturation. 

To conclude, we pose the question if and when the accuracy of gravitational-wave observations will be sufficiently small to provide constraints on the EOS that are tighter than the ones from nuclear theory. From our results, we estimate that the uncertainty $\tilde{\Lambda}$ needs to be of the order of $\Delta\tilde{\Lambda}<300$ to test the chiral EFT prediction in the density range $n_{\rm sat}-2n_{\rm sat}$. Based on the contrast between MM and CSM, we expect that $\Delta\tilde{\Lambda}<100$ is needed to shed light on the possible existence of phase transitions in dense matter.

\begin{acknowledgement}
This work was supported in part by the U.S.~DOE under Grants 
No.~DE-AC52-06NA25396 and DE-FG02-00ER41132, by the LANL LDRD program, and by the National Science Foundation Grant No.~PHY-1430152 (JINA Center for the Evolution of the Elements). 
J.M. was partially supported by the IN2P3 Master Project MAC, "NewCompStar" COST Action MP1304, PHAROS COST Action MP16214.
This research used resources provided by the Los Alamos National
Laboratory Institutional Computing Program, which is supported by the
U.S. Department of Energy National Nuclear Security Administration under Contract No. 89233218CNA000001. Computational resources have been provided by the National Energy Research Scientific Computing Center (NERSC), which is supported by the U.S. Department of Energy, Office of Science, under Contract No. DE-AC02-05CH11231. Computational resources have also been provided by the J\"ulich
Supercomputing Center.
\end{acknowledgement}

\bibliographystyle{epj}
\bibliography{draft}{}

\end{document}